\documentclass[conference]{IEEEtran}

\def\BibTeX{{\rm B\kern-.05em{\sc i\kern-.025em b}\kern-.08emT\kern-.1667em\lower.7ex\hbox{E}\kern-.125emX}}
\usepackage{amsmath}
\usepackage{xspace}
\usepackage{xcolor}
\usepackage{subcaption}
\usepackage{graphicx}
\usepackage{multirow}
\usepackage{wasysym}
\usepackage{makecell}
\usepackage{nicefrac}
\usepackage{fontawesome}
\usepackage{csquotes}

\usepackage[export]{adjustbox}
\usepackage{booktabs}
\usepackage{
  color,
  float,
  epsfig,
  wrapfig
}
\usepackage{arydshln}
\usepackage{hyperref}
\usepackage[capitalise, nameinlink]{cleveref}
\Crefname{equation}{Eq.}{Eqs.}
\Crefname{figure}{Fig.}{Figs.}
\Crefname{tabular}{Tab.}{Tabs.}
\Crefname{section}{Sect.}{Sects.}

\newcommand{\sysnamelong}{DNNShield~}
\newcommand{\sysnamelongw}{DNNShield}
\newcommand{\sysname}{DNNShield~}

\newcommand{\etal}{\emph{et~al.}\xspace}
\newcommand{\resneteighteen}{\mbox{ResNet-18}\xspace}
\newcommand{\cifar}{\mbox{CIFAR-10}\xspace}
\newcommand{\mnist}{\mbox{MNIST}\xspace}
\newcommand{\gtsrb}{\mbox{GTSRB}\xspace}
\newcommand{\naive}{na\"ive }

\newcommand{\setword}[2]{%
  \phantomsection
  #1\def\@currentlabel{\unexpanded{#1}}\label{#2}%
}

\ifCLASSINFOpdf
\else
\fi

\begin{document}
		%
		\title{DNNShield: Embedding Identifiers for Deep Neural Network Ownership Verification}

		\author{\IEEEauthorblockN{Jasper Stang}
			\IEEEauthorblockA{University of Würzburg\\
				jasper.stang@uni-wuerzburg.de}
			\and
			\IEEEauthorblockN{Torsten Krauß}
			\IEEEauthorblockA{University of Würzburg\\
				torsten.krauss@uni-wuerzburg.de}
			\and
			\IEEEauthorblockN{Alexandra Dmitrienko}
			\IEEEauthorblockA{University of Würzburg\\
				alexandra.dmitrienko@uni-wuerzburg.de}}

		\maketitle

		The surge in popularity of machine learning (ML) has driven significant investments in training Deep Neural Networks (DNNs). However, these models that require resource-intensive training are vulnerable to theft and unauthorized use. This paper addresses this challenge by introducing \sysnamelongw, a novel approach for DNN protection that integrates seamlessly before training. \sysnamelong embeds unique identifiers within the model architecture using specialized protection layers. These layers enable secure training and deployment while offering high resilience against various attacks, including fine-tuning, pruning, and adaptive adversarial attacks. Notably, our approach achieves this security with minimal performance and computational overhead (less than 5\% runtime increase). We validate the effectiveness and efficiency of \sysnamelong through extensive evaluations across three datasets and four model architectures. This practical solution empowers developers to protect their DNNs and intellectual property rights.
		

		%

		\section{Introduction}
\label{sec:intro}
Machine learning (ML) has become ubiquitous in our daily lives, driving significant advancements in fields, e.g., speech recognition~\cite{dhanjal2023comprehensive}, object detection~\cite{liu2020deep,krizhevsky2012imagenet,resnet, iandola2016squeezenet}, natural language processing~\cite{collobert2011natural}, and predictive analysis~\cite{benevento2023towards}. During model training, ML extracts generalized patterns from training data. Those patterns, represented by model parameters, are then used to generate predictions for unseen data. 
Recently, larger ML models, so-called Deep Neural Networks (DNN), with larger learning capacities providing more accurate predictions emerged. However, training DNNs necessitates extensive computational resources~\cite{thompson2022computational} and leverages proprietary datasets. Hence, DNNs are valuable assets for model creators, requiring intellectual property (IP) protection measures. Related IP protection methods are Watermarking and Passporting.

\vspace{0.1cm}
\noindent \textbf{Existing DNN IP Protection Methods.} Watermarking~\cite{uchida2017embedding, li2021spreadtransform, tartaglione2021delving, wang2020watermarkingbackpropagation, wang2021riga, rouhani2019deepsigns,lemerrer2019frontierstitching, adi2018turning, zhang2018protecting, zheng2019blindwatermark, guo2018watermarking, jia2021entangledwm} involves embedding stealthy but identifiable markers as reference watermarks within the model during training. The watermarks (identifiers) can be extracted from a previously marked model using a confidential method (also referred to as key). The extracted identifiers serve as evidence of ownership. In case of suspected model copyright infringement, the watermark extracted from the suspected model (using the secret key) can be compared to the model's reference watermark, proving the copyright infringement. However, many DNN watermarking techniques~\cite{uchida2017embedding, li2021spreadtransform, tartaglione2021delving, wang2020watermarkingbackpropagation, wang2021riga, rouhani2019deepsigns,lemerrer2019frontierstitching, zhang2018protecting, zheng2019blindwatermark, guo2018watermarking, jia2021entangledwm} require key secrecy within their threat model. If ownership is proven by disclosing the extraction key and watermark publicly, subsequent ownership watermark verification is not legitimate because the key is now accessible to anyone who observed the initial extraction. Another method~\cite{adi2018turning} does not require key disclosure but relies on manipulating the dataset, which negatively impacts model performance.

\vspace{0.1cm}
\noindent Passporting~\cite{PassportingInitial, PassportingRevisedVersion} was also proposed, which embeds extra trainable \enquote{Passporting} layers within the model architecture, essentially entangling model and Passporting parameters. Removing the trained Passporting layers degrades the model's prediction performance. Therefore, a model can only be efficiently utilized with the untouched Passporting layers which serve as a means for model identification. However, existing Passporting schemes have four drawbacks: First, the trainable Passporting layers introduce additional learning overhead. Secondly, it has been shown to be prone to attacks~\cite{PassportingAttack}, where the Passporting layers are substituted with new layers that mimic their functionality. Third, the proposed method is limited to being used in conjunction with convolutional layers, thus, making it applicable to only a limited set of model architectures. Finally, the system does not adequately protect unfinished training stages with decent prediction performance, known as checkpoints, as the Passporting layers (identifier) are continuously modified throughout the training process. This presents a significant security risk for models with lengthy training durations, as intermediate checkpoints are left unprotected.

\vspace{0.1cm}
\noindent Summarized, Watermarking fundamentally relies on key secrecy allowing only for a single public ownership verification or manipulates the dataset which degrades model performance. Alternatively, Passporting introduces additional training overhead, is vulnerable to adversarial attacks, is restricted to a single type of layer, and cannot protect model checkpoints.

\vspace{0.1cm}
\noindent \textbf{Approach.} 
To address the limitations of existing DNN IP protection methods, we propose \sysnamelong~, a novel approach that integrates publicly known protection layers into the model architecture. The layers are untrainable, and hence remain unchanged during training. The approach seamlessly integrates two types of Protection layers into the model architecture, namely Hadamard and Permutation layers allowing for usage in conjunction with both convolutional and linear layers. Similar to Passporting, the idea behind protection layers is that the data flow through the model is altered, such that the regular model layers are entangled with the Protection layers. As a consequence, an adversary is faced with the dilemma of either leaving the Protection layers unchanged, which allows for reliable model ownership verification or manipulating the Protection layers, resulting in a significant reduction in model performance due to the entanglement of Protection layers with the regular model parameters.

\vspace{0.1cm}
\noindent\textbf{Contributions.} In particular, we make the following contributions:
\begin{itemize}
    \item We propose \sysname, a novel DNN protection method that eliminates the reliance on secret keys, allowing for repeated public ownership claims, while maintaining negligible impact on model performance and training time. 
    \item \sysname~relies on integrating untrainable protection layers into the model architecture, that enable ownership claims. The layers result in negligible training overhead as \sysname~does not introduce additional trainable parameters. Further, as the layers do not change during training intermediate checkpoints and the final model are secured.
    \item We present two types of untrainable Protection layers, namely Hadamard and Permutation layers, that can be seamlessly integrated into various architectures. The layers can be used in conjunction with convolutional or linear layers, hence \sysname~is applicable to most architectures. 
    \item We evaluate \sysname in diverse application scenarios, in particular the approach is evaluated with four model architectures, namely ResNet-18~\cite{resnet}, a Convolutional, a Fully Connected model, and a Vision Transformer~\cite{visiontransformer}. Further, it is evaluated on three datasets: MNIST~\cite{mnist}, CIFAR-10~\cite{cifar}, and GTSRB~\cite{gtsrb}. Moreover, the robustness against third-party manipulation including adaptive adversarial attacks is shown. \sysname~ has negligible impact on the model performance and a runtime overhead of less than 5\% for Hadamard layers in our experiments.
\end{itemize}

\vspace{0.1cm}
\noindent 
In summary, this work introduces a novel and efficient DNN IP protection method, offering a robust defense against various attacks. \sysname~adds novel protection layers into the architecture, that do not rely on secrecy and are untrainable, hence, do not introduce training overhead. As the layers remain unchanged during training, both, intermediate checkpoints and the final model, are secured. Further, they can be used in conjunction with convolutional or linear layers, making them applicable to most modern model architectures. Overall, \sysname~addresses the limitations of existing DNN IP protection approaches while providing an intuitive and robust ownership verification.

\vspace{0.1cm}
\noindent \textbf{Outline.} We provide background information in~\cref{sec:background} and depict the considered scenario and threat model in ~\cref{sec:problemstatement}.~\cref{sec:approach} details the approach of~\sysname~ followed by a security analysis in~\cref{sec:security_analysis}. The evaluation results are reported in~\cref{sec:eval} and~\cref{sec:discussion} elaborates on additional considerations. Finally, related works are discussed in~\cref{sec:relatedwork} before we draw a conclusion in~\cref{sec:con}.

\section{Background}
\label{sec:background}
Below, we provide information on data representations, DNN layers, and important metrics necessary for understanding our approach, as well as common attacks on DNN IP protection.

\subsection{Data Representations}
\label{sec:background:matrices}
Matrices are fundamental components in ML used to represent the data on which models are trained and run. Matrices have dimensionalities ranging from one (referred to as a vector) to multiple dimensions. For image data, the representation usually adheres to a three-dimensional arrangement: $c$, $w$, and $h$, where $c$ initially indicates the three color channels red, green, and blue, while $w$ and $h$ represent the width and height of the image. Certain layers, such as convolutional layers, introduce additional channels, potentially expanding the representation beyond the initial three channels. Unlike convolutional layers, Fully-Connected layers typically abstract away the spatial dimensions, aligning the input along the first dimension. This knowledge is crucial as our approach operates on matrices.

\subsection{Neural Network Layers}
\label{sec:background:nnfundamentals}
To comprehend our approach, it is essential to have a good understanding of certain model layers. Specifically, our approach works in conjunction with linear and convolutional layers. Hence, it is crucial to understand their inner workings.

\vspace{0.1cm}
\noindent \textbf{Fully-Connected Layer.} Those layers are commonly used in most model architectures~\cite{resnet, krizhevsky2012imagenet, iandola2016squeezenet} and perform the calculation $y = x \cdot w^T + b$. The learnable parameters, namely the weights and bias matrices, are represented by $w$ and $b$, while $x$ and $y$ denote the input and output matrices and $\cdot$ is the dot product operation. The transposed operation ($T$) flips the matrix over its diagonal.

\vspace{0.1cm}
\noindent \textbf{Convolutional Layer.}
Fully-Connected (FC) layers, which have a large number of parameters and perform computationally intensive dot product calculations, have limited applicability to image processing. To overcome this challenge, convolutional layers have emerged as an alternative, especially in DNN architectures such as ResNet~\cite{resnet}. Unlike FC layers, which process information over the input simultaneously, convolutional layers analyze the input data in a fragmented and sequential manner. They operate on small receptive fields, which are essentially windows that scan the input image. As the receptive field moves across the image, a filter, also called kernel, which is a small matrix of trainable weights, is applied to each receptive field, effectively extracting relevant features from the data. This sequential approach allows convolutional layers to capture spatial relationships and patterns efficiently. The underlying mathematical method, called cross-correlation, can be used to process multiple input values and produce a more compact output. The number of trainable parameters is significantly reduced as the kernel only contains a fraction of the weights of a FC layer. The input vector, $x$, undergoes a cross-correlation operation denoted by $*$. The resulting output vector, $y$, is calculated using $y = (x * f)$, where $f$ denotes the filter (kernel) used by the convolutional layer. In DNNs a single convolutional layer typically incorporates multiple kernels~\cite{resnet, iandola2016squeezenet, krizhevsky2012imagenet}, each producing an output feature map ($w,h$) that contributes to a distinct channel dimension $c$.

\vspace{0.1cm}
\noindent \textbf{Element-Wise Multiplication Layer.} The element-wise multiplication~\cite{pytorchMulDocu} takes a matrix and outputs a new matrix with identical dimensions, performing element-wise multiplication for each value. The layer calculates $y = x \circ k$. Here, $x$ refers to the input matrix and $k$ refers to the matrix by which $x$ is multiplied. The symbol $\circ$ denotes the element-wise multiplication. Our approach utilizes those layers to embed unique identifiers into the model architecture.

\subsection{Metrics} \label{sec:background:metrics}
In the following, we first introduce cosine similarity, which is used for ownership verification. Then, we introduce the accuracy metric used to evaluate the performance of DNNs.

\vspace{0.1cm}
\noindent 
\textbf{Cosine Similarity.}
\label{sec:background:cosinesim}
Cosine similarity is a fundamental metric for analyzing data, measuring the similarity between two vectors. It is used, for instance, to measure similarity of ML models for backdoor detection~\cite{mesas,crowdguard}. It is calculated by dividing the dot product of two vectors by the product of their magnitudes. Essentially, the alignment of two vectors within an inner product space are measured which yields values that range from -1 to 1. The formula to calculate the cosine similarity is \mbox{$cos = \frac{x_1 \cdot x_2}{\text{max(} \Vert x_1 \Vert_2 \cdot \Vert x_2 \Vert_2 , \epsilon \text{)}}$}, with $x_1$ and $x_2$ being two vectors. If the cosine similarity value is 1, this implies that the vectors are identical and perfectly aligned. Conversely, if the value is -1, the vectors point in opposite directions, indicating complete dissimilarity. Additionally, if the cosine similarity value equals 0, it means the vectors are orthogonal and lack directional alignment. Our approach is based on measuring the similarity between the parameters of two untrainable (fixed parameters) protection layers to determine if one model closely resembles the other.

\vspace{0.1cm}
\noindent \textbf{Model Performance Metrics.}
\label{sec:background:accuracy}
The accuracy metric assesses a model's predictive capabilities, essentially expressing the ratio of correct predictions for a set of input samples to the total number of samples. A dataset is usually divided into training and testing sets to train a DNN model. The training set serves as the basis for model training, while the testing set is utilized to evaluate the model's performance (accuracy) against new and unseen data. This metric is utilized in our approach to determine the performance of a model and yields an accuracy result from 0\% to 100\%. The formula for accuracy is \mbox{$acc = \frac{\text{True Predictions}}{\text{All Predictions}}$}.

\subsection{DNN IP Protection Attacks}
\label{subsec:background:dnnipprotectionattacks}
Two common attacks on DNN IP protection methods are presented below: Fine-Tuning~\cite{tajbakhsh2016finetuning,simonyan2015finetuning} and Pruning~\cite{han2015pruning}. The effectiveness of both attacks on our approach will be evaluated in ~\cref{sec:eval}.

\vspace{0.1cm}
\noindent \textbf{Fine-Tuning.} In Fine-Tuning~\cite{tajbakhsh2016finetuning,simonyan2015finetuning}, the objective is to remove identifiers by continuing model training on a dataset that is comparable to the initial training dataset, assuming familiarity with the training process and hyperparameters to modify the model accordingly~\cite{tajbakhsh2016finetuning,simonyan2015finetuning}. Typically, in Fine-Tuning scenarios, a much smaller dataset than the one used for training is utilized with the goal of minimizing adjustments to the already learned features. Instead, only minor modifications based on the new dataset are desired to attain high accuracy. The learning rate (ratio how quickly a model adjusts its parameters to improve its performance) is commonly lowered in Fine-Tuning~\cite{simonyan2015finetuning, tajbakhsh2016finetuning} to preserve the learned features.

\vspace{0.1cm}
\noindent \textbf{Pruning.} Pruning~\cite{han2015pruning} is a technique used to decrease the size of DNNs for deployment in resource-limited environments. By strategically removing model parameters, Pruning can be employed to remove an embedded identifier while maintaining acceptable model performance (cf. \cref{sec:background:accuracy}) given that adversaries can arbitrarily modify parameters. Pruning entails selectively removing a predetermined percentage of parameters, referred to as Pruning level. Usually, in model Pruning, the focus is on removing the values with the lowest absolute values, as these parameters are deemed to have the least contribution on the overall performance of the model.

\vspace{0.1cm}
\noindent In the following, we will elaborate on the requirements and the threat model for a DNN IP protection method.

\section{Requirement Analysis}
\label{sec:problemstatement}
\vspace{0.1cm}
\noindent \textbf{Motivation.}
We aim to develop a novel ownership verification method for ML models that addresses the limitations of existing approaches. The method must be robust against adversarial modifications, even if the verification process is publicly known. Furthermore, it should not rely on key secrecy or introduce additional parameters to be trained. For extended DNN training periods, intermediate model versions with decent prediction performance are stored as checkpoints. These checkpoints can be stolen and misused, so the novel technique should maintain the same identifier used for verification throughout the entire training process, ensuring that even model checkpoints are secure.

\vspace{0.1cm}
\noindent \textbf{Considered Scenario.}
In our considered scenario, a data owner develops a proprietary ML model trained on a private dataset. The owner should be able to incorporate a unique non-secret identifier into the model, essentially serving as a distinct signature for ownership verification, which we call (non-secret) key or identifier interchangeably. After training, the model is deployed, either in a cloud environment or by selling it to others. However, the model is subsequently misused, e.g., stolen or illegally distributed and put into production by a third-party. It should be possible to identify the model by analyzing the embedded key, regardless of any modifications, such as Fine-Tuning (cf.~\cref{subsec:background:dnnipprotectionattacks}), Pruning (cf. \cref{subsec:background:dnnipprotectionattacks}), or key removal attempts, made by unauthorized third-parties. If the key remains unchanged, the owner can establish legitimate ownership and take action to claim their intellectual property.

\vspace{0.1cm}
\noindent \textbf{Threat Model.}
The attacker has complete knowledge of the model and its architecture, including parameters. This degree of access is known as white-box access~\cite{whitebox}. Furthermore, the adversary can modify the model parameters and architecture, an extremely powerful scenario that allows the attacker to change the model arbitrarily. This level of access also allows the adversary to partition the model at certain layers and use only parts of the model. In addition, similar attacks can be launched on the ownership verification method as those employed on watermarking approaches~\cite{guo2018watermarking, chang2005watermarking, zhang2020modelwatermarking}. For instance, the attacker might modify the model by Fine-Tuning (cf. \cref{subsec:background:dnnipprotectionattacks}) its parameters on a small subset of the training data. Furthermore, an adversary could fine-tune the model to fit a different dataset, e.g., with different output classes.
Moreover, the attacker can launch Pruning (cf. \cref{subsec:background:dnnipprotectionattacks}) attacks that entail eliminating specific connections among parameters within the model. However, the attacker does not have access to the original training data used to train the model, otherwise the attacker could train his own model instead. Furthermore, the attacker cannot tamper with the training and protection process, nor with the training data itself, because the training and protection are performed entirely on the model creator's side, without adversarial influence.

\vspace{0.1cm}
\noindent \emph{Adaptive Adversary.} An adversary who is aware of the protection technique in place, may seek to manipulate or remove it. An adaptive adversary can use arbitrary techniques, leveraging the capabilities defined in the threat model, to remove the identifier. He could launch attacks specifically tailored to the protection method, adapting and extending known adaptive adversary scenarios from the Watermarking and Passporting domains. In general, the goal of the adaptive adversary would be to remove or replace the identifier, e.g., insert an arbitrary identifier. Removing the existing identifier would prevent the model creator from claiming ownership, while replacing it would give the adversary the ability to claim ownership. Specific attack scenarios are discussed in \cref{sec:security_analysis} and \cref{sec:eval}.

\vspace{0.1cm}
\noindent \textbf{Objectives.}
Ideally, protected models should preserve their functionality and allow for unrestricted inference; however, the primary requirement is to maintain a transparent and robust identifier that can withstand attacks. Even as adversaries who have stolen the model adapt it for their own purposes, it is essential to preserve the identifier as evidence of copyright infringement in a legal context. By incorporating unique identifiers, the protection technique aims to secure the IP contained in models while preserving the functionality and performance of the model. Protection methods for DNNs must address the following six fundamental challenges. 

\noindent \setword{\emph{1. Transparency.}}{req:r6:transparency} The ownership verification should be transparent, in particular, it should not rely on the secrecy of the used key. This entails the ability to perform multiple ownership claims.

\noindent\setword{\emph{2. Robustness.}}{req:r1:robustness} The protection technique should be resilient to adversarial attacks to prevent tampering or compromise. Specifically, the model’s distinctive identifier must remain discernible, even after arbitrary adversarial attack attempts, such as Fine-Tuning~\cite{tajbakhsh2016finetuning,simonyan2015finetuning} or Pruning~\cite{han2015pruning} as described above.

\noindent\setword{\emph{3. Fidelity.}}{req:r2:fidelity} The protection should have negligible performance impact, e.g., accuracy should be upheld to levels of unprotected models.

\noindent\setword{\emph{4. Efficiency.}}{req:r3:efficiency} The technique is expected to produce minimal overhead related to complexity and latency, e.g., significant training duration overhead should not occur for protected models.

\noindent\setword{\emph{5. Reliability.}}{req:r4:reliability} The protection process should ensure that unprotected models do not generate false positives, e.g., an unprotected model should not be identified as a specific protected one, while identifying protected models accurately.

\noindent\setword{\emph{6. Generalizability.}}{req:r5:generalizability} The method should be versatile and adaptable to various datasets and model architectures. The flexibility ensures smooth integration into multiple ML architectures, making it valuable across various applications and scenarios.

\vspace{0.1cm}
\noindent In the following section we present our approach that solves the aforementioned challenges and fulfills the objectives.

\section{\sysname~ Design}
\label{sec:approach}
To protect DNNs from unauthorized access, we propose \sysnamelongw, a method that facilitates reliable model identification after deployment by embedding unique identification layers that do not require secrecy and, thus, allow for repeated ownership verification of models. The method integrates non-secret and untrainable protection layers (referred to as locks) with fixed parameters (keys) well-distributed into the model architecture before training, without disrupting functionality or increasing training complexity. \sysname~is superior to watermarking schemes~\cite{uchida2017embedding, li2021spreadtransform, tartaglione2021delving, wang2020watermarkingbackpropagation, wang2021riga, rouhani2019deepsigns, lemerrer2019frontierstitching, adi2018turning, zhang2018protecting, zhang2020modelwatermarking, zheng2019blindwatermark, guo2018watermarking, jia2021entangledwm} because it does not require a secret key for ownership verification and, thus, allow for multiple verifications. The core novelty of~\sysname~lies in the design of novel protection layers. Contrary to existing Passporting~\cite{PassportingRevisedVersion,PassportingInitial} solutions,~\sysname~has three advantages: The protection layers do not require training, which avoids overhead. Further, our approach is not limited to models that use convolutional layers and can also secure models that only use linear layers. Moreover, \sysname~can secure the model at any stage of the training process, including model checkpoints, with the same key. This is important as models become larger and their training time increases.

\vspace{0.1cm}
\noindent Each protection layer has a unique key (parameters) that is not changed during training. The key determines how the protection layer input, which is the output of the previous layer, is altered. For instance, the key defines how the data is scaled. Therefore, static operations on the output of the previous layer are applied and, hence, the behavior of subsequent layers is influenced. The altered data introduced by protection layers ensures that the model parameters of subsequent layers are entangled with the specific protection layer key. Therefore, protection layers cause a significant model performance reduction if they are removed or their keys altered, rendering the model inoperable. In particular, the data processed by the manipulated model would deviate from the expected pattern, resulting in unexpected inputs to the layers following the protection layers. We name this characteristic of protection layers \enquote{protection property}. Effectively, adversaries are compelled to maintain the integrity of the protection layers by using identical or very similar keys, facilitating ownership verification of legitimate owners by analyzing the used key. Similar to Passporting~\cite{PassportingInitial, PassportingRevisedVersion} ownership verification schemes, the keys used by the suspected model are compared to the original keys used by the model owner, and if there is a high degree of similarity, model ownership can be claimed. Therefore, the unique keys of the protection layers can be published along with the model and then serve as a unique identifier, allowing for multiple ownership claims. This fulfills the \hyperref[req:r6:transparency]{Transparency} requirement from~\cref{sec:problemstatement}. To summarize, the protected model can be deployed and the lock and key's persistence allows the model to be identified. If adversaries attempt to remove the protection, the model performance is significantly degraded, rendering the model inoperable. In the following, we present the steps for protecting a model using \sysname.

\begin{figure}[t]
\includegraphics[width=0.5\textwidth]{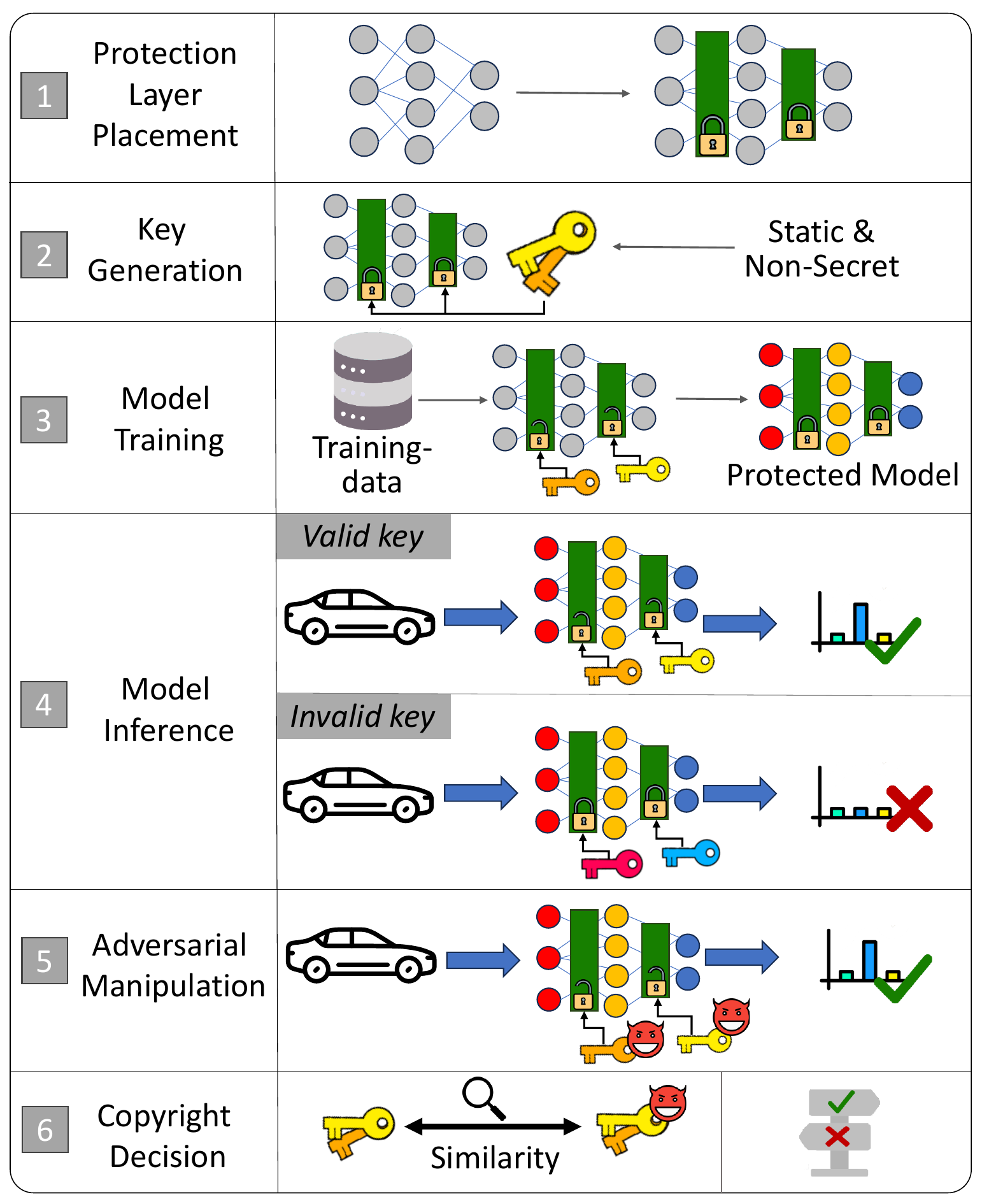}
\centering
\caption{Overview of the Approach.}
\label{fig:approachOverview}
\end{figure}

\subsection{General Approach}
Below, we present the life cycle of the \sysname~method consisting of six stages. All stages are depicted in \cref{fig:approachOverview}.

\vspace{0.1cm}
\noindent \textbf{Protection Layer Placement.} In the initial stage, visualized as step 1 in \cref{fig:approachOverview}, protection layers are incorporated at strategic points in the model architecture after unprotected layers, such as Convolutional or Fully-Connected (FC) layers (cf. \cref{sec:background}). We elaborate on the exact placement and amount of protection layers in ~\cref{subsec:protectionLayerPlacement}.

The layers do not require training and, therefore, do not introduce additional training overhead. A protection layer modifies the output of the previous layer while maintaining the same output dimensionality, making it compatible with the following layer. This allows integration into existing model architectures at any point.\footnote{Otherwise, incorporating an extra dimension adapter layer would be required to adjust the output to the appropriate dimensions, which is not desired for easy integration and achieving minimal training overhead.} Therefore, the requirements \hyperref[req:r3:efficiency]{Efficiency} and \hyperref[req:r5:generalizability]{Generalizability} from~\cref{sec:problemstatement} are addressed. The protection layer and its introduced data alteration become an integral component of the model since in DNNs the following layers rely on the output of the previous ones. Protection layers protect the parts of the model that follow them. Therefore, integrating multiple protection layers into a single model creates a more comprehensive defense by increasing the protection coverage.

\vspace{0.1cm}
\noindent \textbf{Key Generation.} Each protection layer is associated with a unique non-secret key in the form of static parameters that define how the data is altered. The protection layer serves as a gatekeeper, guaranteeing proper functionality only when the correct key is provided (referred to as protection property). Therefore, the second stage (step 2 in \cref{fig:approachOverview}) involves defining a unique non-secret key for each of the previously defined protection layers. The composition, quantity, and length of the keys, which are the protection layer static parameters, vary depending on the protection layer type and the model architecture. In \cref{subsec:protectionLayerInstantiations} we will outline details of the key generation process. Once generated, the keys remain unchanged, including during model training and inference.

\vspace{0.1cm}
\noindent \textbf{Model Training.} Next, as shown in step 3 in \cref{fig:approachOverview}, the model is trained with the protection layers and corresponding keys in place. The training process and the model creator's data is not manipulated. A model adjusts its parameters based on the specific data alterations from the protection layers defined by the used key. Thus, the parameters of the model are entangled with the keys. In the subsequent step, the trained and protected model along with the keys is then deployed, e.g., to the end-user or a web service.

\vspace{0.1cm}
\noindent \textbf{Model Inference.} Model inference (step 4 in \cref{fig:approachOverview}) uses the keys defined in step 2 to unlock the protection layers. If the correct keys are provided to the protection layers within the model architecture, the data alteration will be identical to the one during training and the model will exhibit high performance. However, if the keys are not correct, e.g., manipulated, the performance will significantly decrease. This is caused by the protection layers not producing the expected data, causing subsequent layers that depend on the data produced by the protection layers to fail. In essence, a mismatch between the data from the protection layers and the learned parameters from the following layers occurs. Therefore, an adversary is compelled to use the keys defined in the second step; otherwise, the model becomes useless in terms of exhibited accuracy.

\vspace{0.1cm}
\noindent \textbf{Adversarial Manipulation.} In step 5 of \cref{fig:approachOverview}, an adversary obtains and misuses the model and is capable of performing arbitrary manipulations as defined in the threat model in~\cref{sec:problemstatement}. For instance, attempts to manipulate the utilized keys, the parameters of the model or its architecture could be performed (cf.~\cref{sec:problemstatement}). The adversary's goal is to unlock the protection layers with different self-defined keys or remove them while maintaining the model's accuracy. This allows for unrestricted use of the model as copyright claims can no longer be made.

\vspace{0.1cm}
\noindent \textbf{Verification.} In step 6 of \cref{fig:approachOverview}, all keys of a suspected model are analyzed. A comparison is drawn between the original keys from all protection layers and those from the suspected model. Our approach employs metrics that are suitable for the novel version of protection layers. Copyright infringement will be evident if significant similarity is detected. We will elaborate on how to reliably measure the similarity of different keys in~\cref{subsec:protectionLayerInstantiations}, by utilizing the cosine similarity and a newly introduced metric.

\subsection{Protection Layer Instantiations}
\label{subsec:protectionLayerInstantiations}
In this section, we present two specific instantiations of protection layers. One is based on element-wise multiplication called \enquote{Hadamard layer}, while another one relies on shifting the order of outputs called \enquote{Permutation layer}.

\begin{figure}[t]
\includegraphics[width=0.45\textwidth]{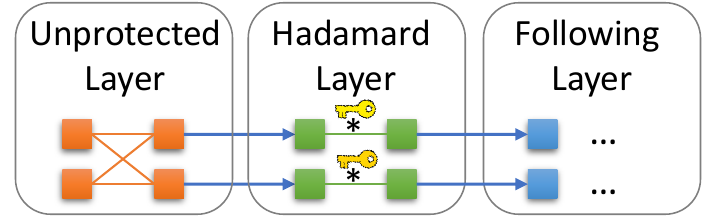}
\centering
\caption{The Hadamard layer Building Block.}
\label{fig:hadamardbuildingblock}
\end{figure}

\vspace{0.1cm}
\noindent\textbf{Hadamard Layer.} Those layers perform an element-wise multiplication (cf. ~\cref{sec:background:nnfundamentals}) between two matrices of equal dimensions. The protection layer can be used in conjunction with any convolutional or FC layer (addressing the \hyperref[req:r5:generalizability]{Generalizability} requirement of~\cref{sec:problemstatement}). As shown in~\cref{fig:hadamardbuildingblock}, the protection property is implemented by using the output values of the preceding layer as input for the protection layer and multiplying each of them with the corresponding key value, visualized by the green boxes and the keys in the center. Thus, the key must have equal dimensionality as the output (usually $c,w,h$ for convolutional layers or $w,h$ for FC layers as explained in~\cref{sec:background:nnfundamentals}) of the preceding layer. In essence, every output value of the preceding layer, undergoes scaling by a different key-defined constant factor. The following layers adjust their functionality to the particular scaling introduced by the Hadamard layer during model training.

\vspace{0.1cm}
\noindent\emph{Robustness.} If the scaling defined by the key deviates significantly during inference, it will result in changes to the output values. As all subsequent layers rely on values produced by the Hadamard layer during training changing the keys will lead to decreased performance. This fulfills the \hyperref[req:r1:robustness]{Robustness} requirement from~\cref{sec:problemstatement}, as the layers cannot handle values that deviate from the known scale. A misalignment in value scaling impacts the entire model, including all subsequent layers, as each layer depends on the output values of the previous layer. We show the functionality and robustness of the Hadamard layer through an empiric evaluation in \cref{sec:eval}.

\vspace{0.1cm}
\noindent\emph{Key Generation.} For every output value in the preceding layer, an element-wise multiplication is executed with a specific value provided by the non-trainable key. As such, the key has to contain one scaling factor for each output value. While there are no general limits on the values within the key, it is important to choose value ranges carefully to prevent gradients from exploding or vanishing~\cite{gradientproblems}. Values that approach zero create particularly small output values and can subsequently lead to vanishing gradients. Similarly, excessively large key values may lead to exploding gradients, and should therefore be avoided as specified by the \hyperref[req:r2:fidelity]{Fidelity} requirement of~\cref{sec:problemstatement}. 
We propose to use randomly selected values that are uniformly distributed within a range of -1.0 to 1.0. Negative and positive values are included in the key to enable changing signs, which, we believe, makes it more difficult to remove protection layers due to the increased entanglement with model parameters. We explore the effects of different key ranges in~\cref{sec:eval}.

\vspace{0.1cm}
\noindent\emph{Copyright Verification.} The keys of the Hadamard layers serve as model identifiers. Therefore, the keys of a suspected model are compared one by one to the original keys. For copyright verification, the keys are represented as flattened vectors and we suggest using cosine similarity (cf. \cref{sec:background:cosinesim}) for comparison. While the specific comparison metric may differ implementation-specific utilizing cosine similarity can provide a reliable measure of similarity, essentially fulfilling the \hyperref[req:r4:reliability]{Reliability} requirement (cf.~\cref{sec:problemstatement}). If there is a high similarity between the original and suspected model keys, it is likely that the entire model or parts of it were copied from the original model, providing evidence of copyright infringement. Similarity calculations can be performed for each Hadamard protection layer key individually, providing a detailed understanding of which model layers were stolen. We establish an insensitive guideline regarding the cosine similarity for ownership verification through an empirical study in \cref{sec:eval}. Furthermore, we empirically show that, despite third-party manipulation post-deployment, the following layer's dependence on a specific value range cannot be eliminated.

\vspace{0.1cm}
\noindent Next, we introduce the Permutation layer, a protection layer that was specially crafted to be used with convolutional layers (addressing the \hyperref[req:r5:generalizability]{Generalizability} requirement from~\cref{sec:problemstatement}). The robustness of the Permutation layer relies on its usage in conjunction with convolutional layers. It could function as an alternative or extension to the Hadamard protection layer.

\begin{figure}
    \centering
    \includegraphics[width=0.8\linewidth]{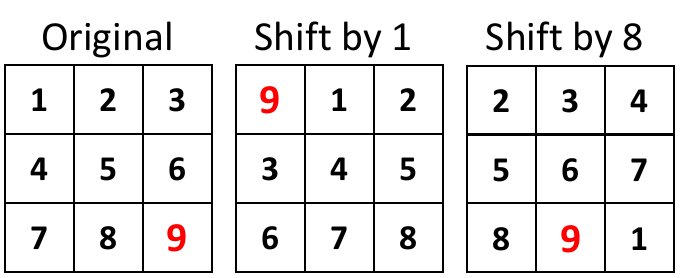}
    \caption{A shift to the right by 1 and 8 positions is shown.}
    \label{fig:aproach:matrixshift}
\end{figure}

\vspace{0.1cm}
\noindent \textbf{Permutation Layer}
The Permutation layer shifts the order of input data, making it an effective mechanism for implementing a protection layer that adheres to the protection property. The intuition behind the layer is to allow the convolution process begin at a unique non-standard starting position, rather than the top-left corner, while still processing the entire image sequentially. The key defines the specific starting position for each kernel. The resulting output retains the dimensions from the unprotected layer and preserves local relationships between adjacent output values while adopting a rearranged order. Recall, that each kernel in the convolutional layer (cf.~\cref{sec:background:nnfundamentals}) has its own output channel. Instead of modifying the starting positions of the kernels, one can achieve the same effect by shifting the outputs of the convolutional layer on a channel-wise basis. An example of a shift from a single channel to the right within a 3x3 matrix is shown in \cref{fig:aproach:matrixshift}. The left matrix indicates the original order, while the middle matrix shows a shift to the right by one position and the right matrix shows a shift by eight positions to the right. The non-trainable and static key must have the same length as the number of output channels (cf. \cref{sec:background:nnfundamentals}) of the preceding convolutional layer to ensure proper operation. During training, subsequent layers will adjust their parameters based on the output order introduced by the Permutation layer.

\vspace{0.1cm}
\noindent\emph{Robustness} 
A deviation from the altered order (defined by the key) of the Permutation layer disrupts the learned features, rendering the following layers incapable of accurate operation, fulfilling the \hyperref[req:r1:robustness]{Robustness} requirement of~\cref{sec:problemstatement} and adhering to the protection property. Random shuffling cannot be performed because convolutional kernels may partially overlap during the convolution operation, effectively operating on the same pixels multiple times. Consequently, the outputs exhibit a sensitivity to their order, meaning randomly shuffling them can significantly impair the model performance as spatial information is lost. To maintain the correct relative positions, the values are shifted (addressing the \hyperref[req:r3:efficiency]{Efficiency} requirement from~\cref{sec:problemstatement}). We confirm the protection effectiveness of this layer through empirical evaluation in \cref{sec:eval}.

\vspace{0.1cm}
\noindent\emph{Key Generation}
Recall, that the number of output channels is dependent on the number of kernels utilized in the convolutional layer (cf.~\cref{sec:background:nnfundamentals}). In the Permutation layer, all outputs from a channel are shifted to the right by a certain key-defined factor. For instance, the values in the first channel will be shifted by three positions, while values from the second channel will be shifted by five positions. Thus, the key consists of a single integer value for each channel of the preceding convolutional layer. The key values are randomly generated between one and the number of output values minus one in the respective channel. This indicates a shift to the right by one position or a shift to the left by one position. The careful selection ensures that the layer shifts at least one position to the right and in each case alters the input's order.

\vspace{0.1cm}
\noindent\emph{Copyright Verification.}
Instead of directly comparing the keys of the Permutation layers as done with Hadamard layers, we examine the outputs of Permutation layers to establish ownership verification. This is done to mitigate adaptive adversary attacks, which will be explained in~\cref{sec:security_analysis}. To achieve this, we first create a synthetic input sample where each value is unique, such as a matrix with entries ranging from 0 to the number of data points as shown in \cref{fig:aproach:matrixshift}. Then, we analyze the outputs of the input sample generated by two Permutation layers. The first Permutation layer is used with the originally employed key, while the second uses the key found in the suspected model. Given the absence of suitable methods to assess the similarity between two shifted outputs, we devised a novel metric. This metric involves calculating the channel-wise number of shifts required to move one of the outputs to the left or right, maximizing the resemblance to the other. Similar to brute-forcing, every possible shift is performed and the output that produces the highest similarity is selected. In particular, the cosine similarity is utilized to measure the similarity between the two outputs.

\vspace{0.1cm} 
\noindent We start by defining $k=\frac{\mathrm{Possible Shifts}}{2}$ as the maximum number of shifts (divided by two as shifting can be done to the left or right side). Next, we define the Permutation Accuracy (PAC) as $\mathrm{PAC} = 1 - \nicefrac{r}{k}$. Here, $r$ specifies the minimum count of shifts that maximized the similarity between both outputs (considering both left and right direction). This metric defines the percentage of similarity between the output values of two Permutation layers in terms of value order. As the output order is defined by the key, essentially, the key's similarities are measured. Thus, ownership can be claimed based on the similarity of the PAC metric.

\subsection{Protection Layer Placement}
\label{subsec:protectionLayerPlacement}
In the following, we elaborate on the placement and number of protection layers in model architectures. Protection layers can be placed arbitrarily, as long as they are preceded by a convolutional or FC layer. Integrating multiple protection layers distributed at different locations provides more comprehensive protection, as each layer protects the parts of the model that follow it.

\vspace{0.1cm}
\noindent DNN Models are often already split into architectural parts as their architecture repeats itself, e.g.~\cite{resnet,krizhevsky2012imagenet,iandola2016squeezenet,mobilenet}. For instance, the ResNet ~\cite{resnet} models are constructed from a number of basic blocks. Similarly, Transformer-based models consist of a number of Transformer blocks~\cite{llama, gpt3}. The amount and exact placement of Protection layers is an insensitive parameter as we show in ~\cref{eval:subsec:generalfunc}. Nevertheless, we propose placing one protection layer inside each model block after the first FC or Convolutional layer, providing comprehensive model protection. The last layer of a model is the output layer, here no Protection layer is added as no layer uses these outputs for further computation.

\vspace{0.1cm}
\noindent For instance, a \resneteighteen is divided into its nine Basic Blocks where each block has roughly the same number of layers as shown in Appendix ~\cref{app:tab:resnet}. The first convolutional layer of each block is preceded by a Protection Layer. Likewise, a Transformer-based model is protected by placing one Protection layer after each FC layer from each Transformer block. Smaller models can also intuitively be split into model parts where each part has roughly the same amount of layers as demonstrated in Appendix ~\cref{tab:app:fcn} and ~\cref{tab:app:cnn}.

\section{Security Analysis}
\label{sec:security_analysis}
In the following, we discuss attacks that merge the protection layer with neighboring layers and attacks that manipulate convolutional kernel patterns to imitate the Permutation layer. Additionally, the feasibility of partitioning the protection layer and the robustness of the PAC similarity metric (cf. ~\cref{subsec:protectionLayerInstantiations}) are discussed.

\vspace{0.1cm}
\noindent \textbf{Merge.}
\begin{figure}[t]
\includegraphics[width=0.4\textwidth]{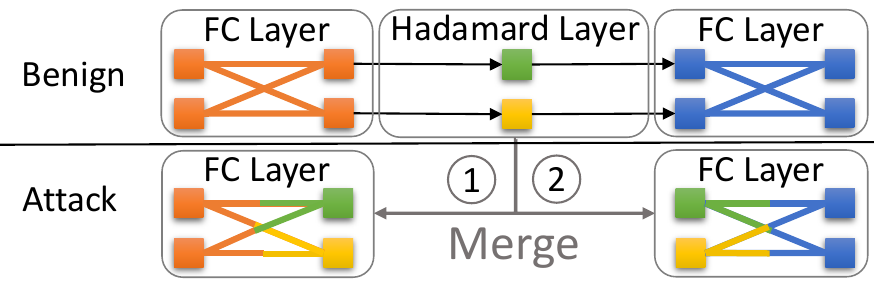}
\centering
\caption{Merge of Hadamard layer into Fully-Connected (FC) layers.}
\label{fig:adaptivemerge}
\end{figure}
If a Hadamard layer is used in conjunction with a Fully-Connected (FC) layer, an adversary could merge the keys of the Hadamard layer with a FC layer (merge attack), as illustrated in ~\cref{fig:adaptivemerge}. The key of the Hadamard layer can either be multiplied with the parameters of the preceding layer (annotated with 1) or with the following FC layer (annotated with 2). Merging one of the FC layers and the Hadamard layer would allow for the removal of the Hadamard layer. We elaborate on the mathematical details of the multiplication in~\cref{app:weightmerge}. However, this attack still leaves identifiable traces that can be used for copyright claims.

\vspace{0.1cm}
\noindent Assume that an adversary performs the merge attack and, therefore, obtains a manipulated (FC) layer that does not require the Hadamard layer for proper outputs anymore, as visualized as the bottom left FC layer or the bottom right FC layer in~\cref{fig:adaptivemerge}. The manipulated layer comprises of parameters that are multiplied by the key of the Hadamard layer. For example, in the lower left corner, the outgoing paths from the first orange neuron are multiplied by the green factor of the Hadamard key. The key of the Hadamard layer can still be extracted from the manipulated layer by reversing the merge attack. The key is obtained by dividing the parameters of the manipulated layer by the parameters of the original FC layer. As visualized in \cref{fig:adaptivemerge}, each neuron has multiple paths and for a simple merge attack each parameter would yield the same key factor. However, in case the adversary manipulates a parameter, all the paths contributing to one output must be considered. Hence, we average all multiplication factors determined by all paths contributing to one output. The division yields a multiplication factor corresponding to the key for each path. The mathematical details of the reversal process are outlined in \cref{app:keyextraction}. After averaging, the obtained values have identical dimensionality as the originally used key and can be easily compared using the cosine similarity metric. Furthermore, we demonstrate through a robustness evaluation in~\cref{subsec:eval:robustness} that this attack combined with Fine-Tuning the resulting manipulated model is also ineffective.

\vspace{0.1cm}
\noindent Convolutional layers cannot incorporate the Hadamard layer's key into the parameters, as convolutions share the same parameters for multiple outputs. Therefore, the merge attack would require multiplying a single parameter with different factors simultaneously, which is not possible. We elaborate on the precise details in \cref{app:cnnmerge}. Thus, the merge attack is infeasible for both FC and convolutional layers.

\vspace{0.1cm}
\noindent \textbf{Convolutional Pattern Modification.}
\begin{figure}[t]
\includegraphics[width=0.35\textwidth]{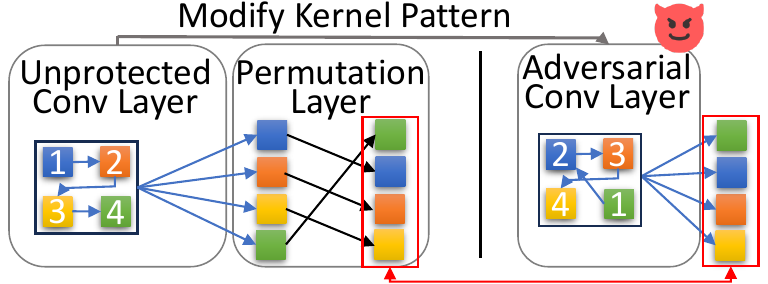}
\centering
\caption{Kernel pattern modification attack.}
\label{fig:adaptiverkernelpos}
\end{figure}
Permutation layers shift output values by a certain factor (cf. ~\cref{subsec:protectionLayerInstantiations}). Rather than modifying the order from the convolutional layer, the same rearrangement could be produced by manipulating the starting position of the convolutional kernel as shown in~\cref{fig:adaptiverkernelpos}. The figure visualizes, that modifying the order in which the convolution is performed on the right-hand side results in the same output order as introduced by the Permutation layer in the center. This attack allows for removal of the Permutation layer. However, ownership can still be proven. This is done by calculating the PAC metric on the output of the adversarial convolutional layer and the output of the original convolutional layer combined with the Permutation layer. As only the order of output values is important, the parameters of the manipulated and original convolution layers are unified, e.g., they have the same parameters. Thus, the order in which the output is rearranged is analyzed for both layers, and ownership can be claimed as the same order is produced, rendering the attack infeasible.

\begin{figure*}
        \captionsetup[subfigure]{aboveskip=-2pt,belowskip=-2pt}
	\centering
	\hfill
        \begin{subfigure}{0.15\linewidth}
		\includegraphics[width=\linewidth]{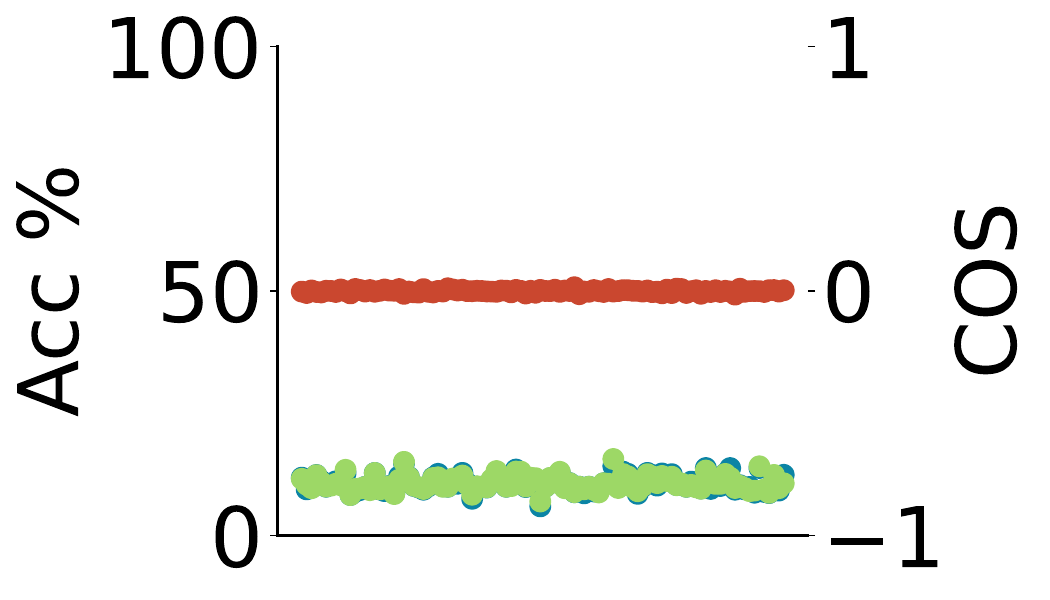}
		\caption{Random keys}
		\label{fig1:subfigA}
	\end{subfigure}
	\begin{subfigure}{0.15\linewidth}
		\includegraphics[width=\linewidth]{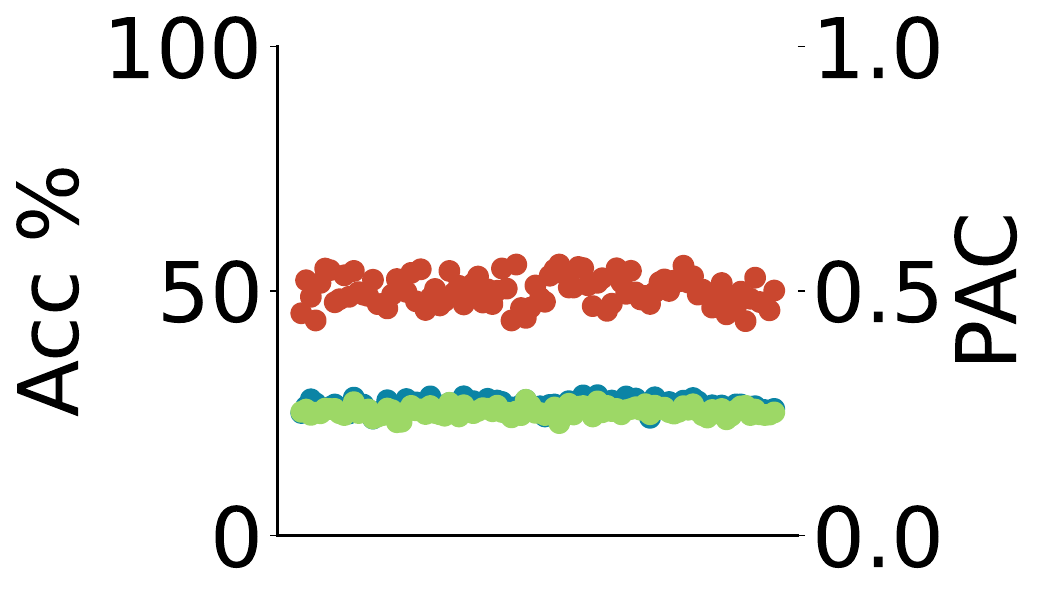}
		\caption{Random keys}
		\label{fig1:subfigB}
	\end{subfigure}
	\hfill
        \begin{subfigure}{0.15\linewidth}
		\includegraphics[width=\linewidth]{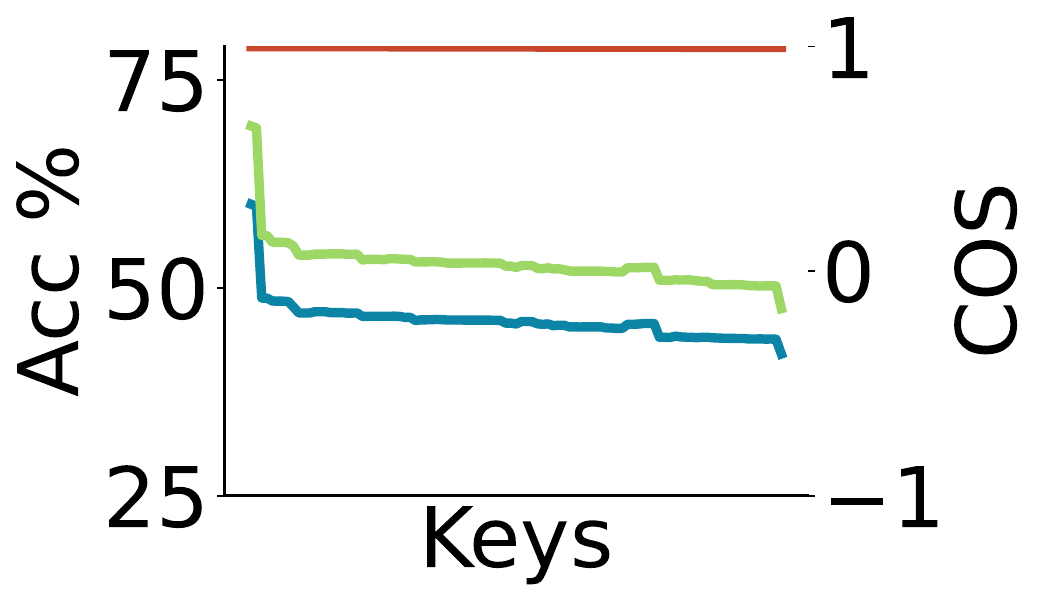}
		\caption{Random value}
		\label{fig1:subfigC}
	\end{subfigure}
	\begin{subfigure}{0.15\linewidth}
		\includegraphics[width=\linewidth]{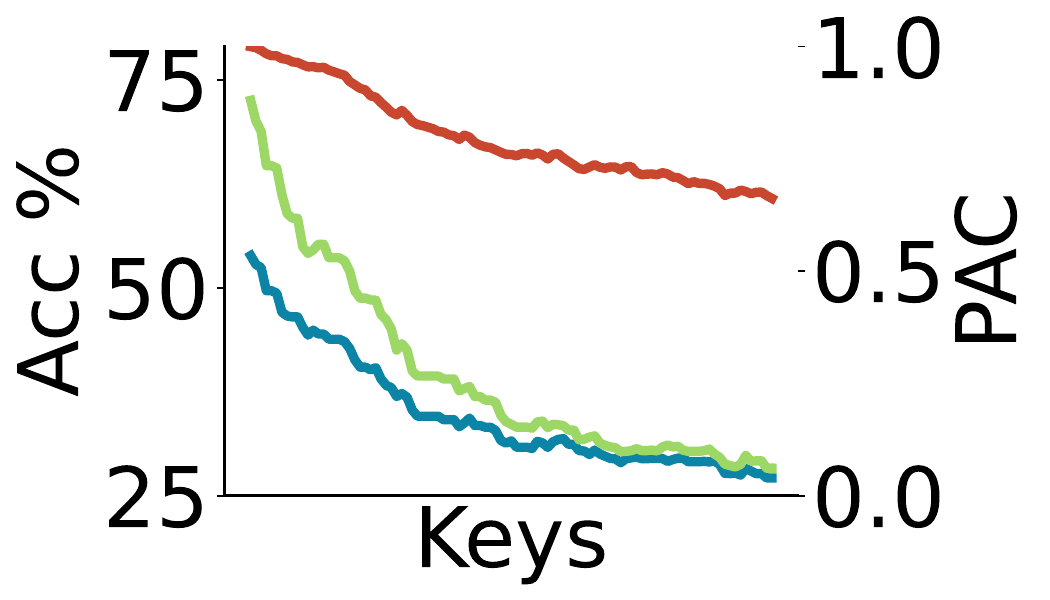}
		\caption{Random value}
		\label{fig1:subfigD}
	\end{subfigure}
        \hfill
        \begin{subfigure}{0.15\linewidth}
            \includegraphics[width=\linewidth]{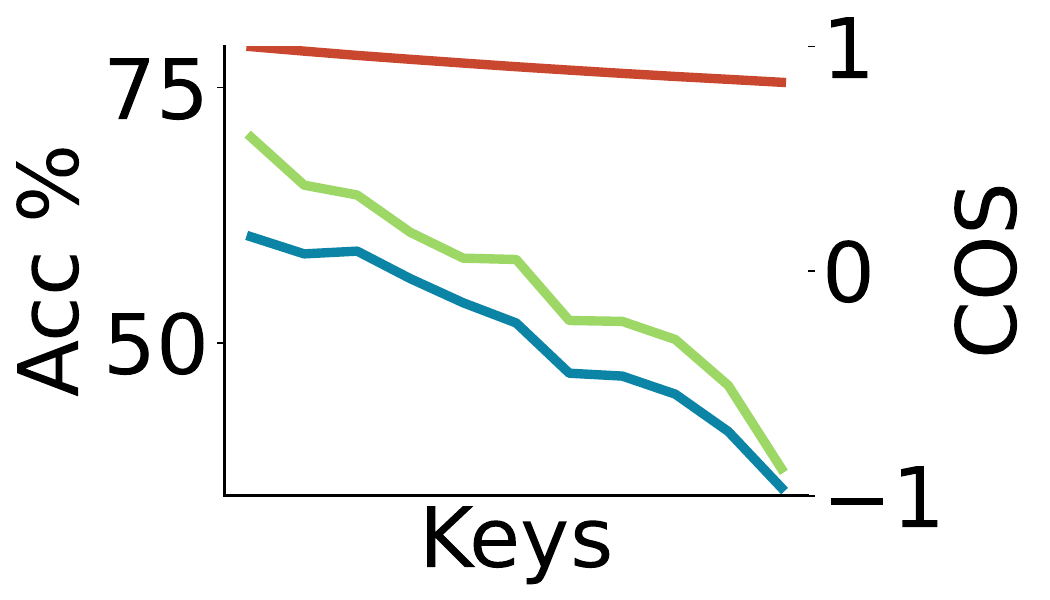}
            \caption{Add noise}
            \label{fig1:subfigE}
        \end{subfigure}
        \begin{subfigure}{0.15\linewidth}
            \includegraphics[width=\linewidth]{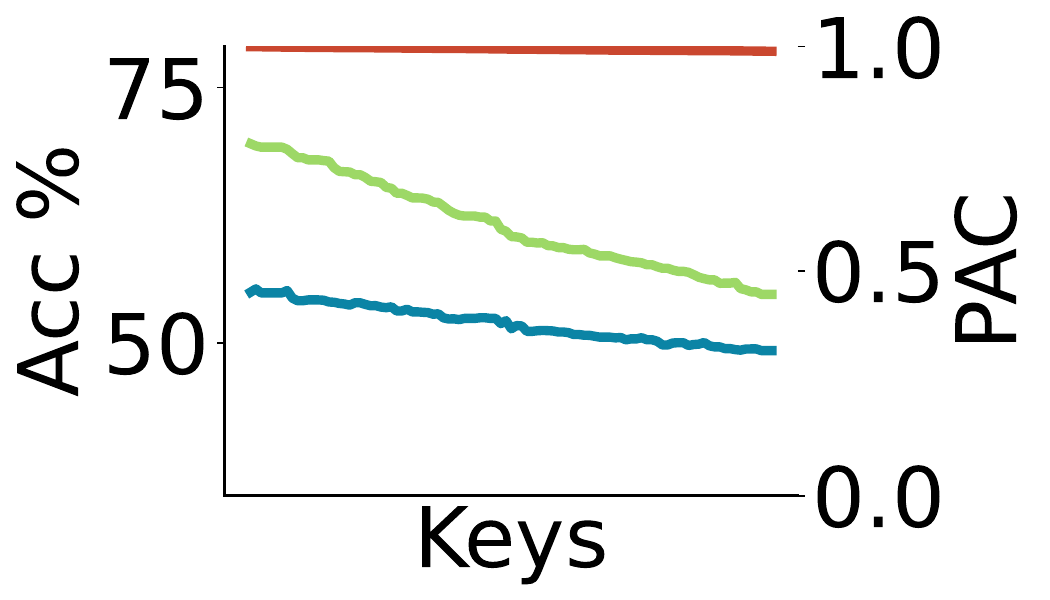}
            \caption{Increment value}
            \label{fig1:subfigF}
        \end{subfigure}
        \hfill
 \caption{CNN robustness protected by Hadamard(a,c,e)/Permutation layers(b,d,f) to key manipulation, e.g. random key, replacing of key value, and, adding noise to key. Red line depicts key similarity, Green and Blue depict train and test accuracy.}
	\label{fig:subfigures}
\end{figure*}

\vspace{0.1cm}
\noindent \textbf{Protection Layer Split.}
A potential attacker may try to bypass the protection mechanism by dividing the protection layers into multiple layers, each performing a part of the overall function. However, the cumulative effect of these layers would replicate the original protection layer. Therefore, it can be identified by examining the architecture, as the layers would need to be sequentially connected. Consequently, the separated layers can be easily reassembled into a single layer, and ownership can be claimed.

\vspace{0.1cm}
\noindent \textbf{Similarity Metric Resilience.}
The proposed PAC metric relies on a minimal distance to assess the similarity between Permutation layer outputs. This prevents an adversary from subverting the PAC similarity metric by introducing a single value that has not been shifted but rather randomly permuted. Even if the injected value disrupts the alignment of the two outputs, making it infeasible to align them with a shifting operation alone, the PAC metric can still determine the number of shifts required to maximize their resemblance. As a result, this renders the attack ineffective.

\section{Evaluation}
\label{sec:eval}
\noindent 
\textbf{Model Architectures and Datasets.}
Our approach was assessed on various model architectures with differing sizes commonly utilized in image classification domains, including a Fully Connected network (FCN), a Convolutional network (CNN), and the \resneteighteen~\cite{resnet} architecture. Further information about these models can be found in \cref{app:modeloverview}. We chose the Vision domain to showcase the effectiveness and robustness of \sysname, aligning with prior research (\cite{uchida2017embedding, li2021spreadtransform, tartaglione2021delving, wang2020watermarkingbackpropagation, rouhani2019deepsigns, lemerrer2019frontierstitching, adi2018turning, zhang2018protecting, zheng2019blindwatermark, guo2018watermarking}), leveraging commonly used datasets that focus on image classification namely \mnist~\cite{mnist}, \cifar~\cite{cifar}, and \gtsrb~\cite{gtsrb}. Throughout the experiments, we varied the model architecture, and training dataset systematically to showcase the \hyperref[req:r5:generalizability]{Generalizability} (cf.~\cref{sec:problemstatement}) of our approach. The models underwent ten epochs of training utilizing the Adam optimizer with a learning rate of \emph{0.001}, unless explicitly stated otherwise. In all experiments, the mean similarity of all keys is reported. The experiments were conducted using PyTorch, a leading Python-based machine learning library~\cite{pytorch, paszke2019pytorch, van1995python}, on a server equipped with 96 processing units, 128GB main memory, and an AMD EPYC 7413 24-Core Processor (64-bit). We accessed an NVIDIA A16 GPU via CUDA~\cite{cuda}, which has four virtual GPUs, each with 16GB of GDDR6 memory.

\subsection{\sysname's Functionality}
\label{eval:subsec:generalfunc}
To showcase the general functionality of \sysname, we employ Hadamard and Permutation protection layers to protect the CNN model (cf. \cref{app:modeloverview}). The comparison results with an unprotected model, as well as the best key ranges, will be evaluated in ~\cref{sec:eval:fidelity}. After training on the \cifar dataset~\cite{cifar}, training accuracies of \emph{69.89\%} for the version with Hadamard layers (H-Model) and \emph{75.77\%} for the version with Permutation layers (P-Model) were achieved, showing that working models can be trained with protection layers in place. The number of values in the key, e.g., the key size of the H-Model was around \emph{3.2\%} compared to all model parameters and around \emph{0.01\%} for the P-Model. First, the functionality of the protection layers is evaluated with three experiments. The goal is to ensure that the protection layers reduce the model accuracy for keys that deviate from the reference keys (as defined by the \hyperref[req:r1:robustness]{Robustness} requirement from~\cref{sec:problemstatement}) which validates our claims that the protection layers adhere to the protection property (cf.~\cref{sec:approach}). These experiments showcase that the keys are entangled with the model parameters and, thus, have an impact on the model performance in case of manipulation. In the first experiment, we establish a baseline by replacing the key with randomly generated false keys and showing which key similarity and model performance is achieved. Next, we show that small deviations in the key's values have an impact on the model accuracy. Further, we show the impact of adding different levels of noise to the key of a protection layer.

\vspace{0.1cm}
\noindent \textbf{Position and Amount of Protection Layers}
To assess the impact of the number and placement of protection layers within a model, all combinations of one, two, three, and four protection layers integrated into the CNN model (cf. Appendix ~\cref{tab:app:cnn}) were evaluated. The results, depicted in ~\cref{tab:protectionlayer:combinations}, show that neither the number of protection layers nor their specific positions within the model seem to affect the performance. In particular, the columns of the table show the different combinations of active protection layers, e.g. Protection Layer 1 to 4. The rows show the mean model performance and its variance during 20 training epochs. It is evident, that the performance does not change significantly for any Protection Layer combination. This confirms, that the placement and quantity of protection layers is an insensitive parameter. To further showcase the position independence, we evaluated placing the Protection layers after the first within each Basic block of the \resneteighteen~\cite{resnet} as shown in ~\cref{app:tab:resnet}. We also place the Protection layer after the second Convolutional layer within each block. Varying the position of the Protection layers in the \resneteighteen~\cite{resnet} architecture did not alter the outcome of performance or robustness experiments.

\begin{table*}
\vspace{0.4cm}
\centering
\resizebox{1.0\textwidth}{!}{
\begin{tabular}{lccccccccccccccc}
\toprule
Combination & \{1\} & \{2\} & \{3\} & \{4\} & \{1,2\} & \{1,3\} & \{1,4\} & \{2,3\} & \{2,4\} & \{3,4\} & \{1,2,3\} & \{1,2,4\} & \{1,3,4\} & \{2,3,4\} & \{1,2,3,4\} \\
\midrule
Mean Accuracy & 98.4 & 97.9 & 98.9 & 98.9 & 97.7 & 98.6 & 98.4 & 97.7 & 97.9 & 98.8 & 97.9 & 97.7 & 98.4 & 97.8 & 97.6 \\
Accuracy Variance & 0.9 & 1.1 & 0.9 & 0.9 & 1.5 & 1.1 & 1.0 & 2.1 & 1.1 & 1.0 & 1.9 & 2.3 & 1.4 & 1.7 & 3.0 \\
\bottomrule
\end{tabular}}
\caption{Different Combination of protection layer P1, P2, P3, and, P4 for the CNN model trained on MNIST for 20 epochs.}
\label{tab:protectionlayer:combinations}
\end{table*}

\vspace{0.1cm}
\noindent \textbf{Model Refinement}
In scenarios involving data concept drift, where fine-tuning the model is required at later stages to adapt it to new data~\cite{conceptdrift}, it is crucial that the model can be refined after it was protected without requiring re-training from scratch. To evaluate the ability to fine-tune a protected model, we conducted an experiment, training a \resneteighteen model on the \gtsrb dataset for 10 epochs, achieving a training accuracy of 86.95\%. The last layer of the model was then replaced to adapt it to the \cifar dataset. Afterward, the model was fine-tuned on the \cifar dataset for an additional ten epochs starting at an accuracy of 71.59\% and ending with 93.5\%. The used keys remained identical throughout the whole process. These results demonstrate the ease of refining existing models on new datasets, as long as the legitimate keys are preserved. Thus, \sysname~enables model refinement after protection, eliminating the need for complete retraining, resulting in significant resource savings.

\vspace{0.1cm}
\noindent \textbf{Key Replacement.} To determine the functionality and validate the protection property of the layers, we evaluated the trained and protected models in a setup where the original key was replaced with 100 different randomly generated false keys. \footnote{The keys were generated in a valid range, e.g., -1.0 to 1.0 for Hadamard layers and 1 to the number of channels minus one for Permutation layers.} We expect the key similarity to be low, however, also the model performance should significantly drop as the wrong key is used. In \cref{fig1:subfigA} the results are shown, we omit the x-axis label due to space reasons. The green dots represent the accuracy on the train set, while the blue dots represent the accuracy on the test set. Note, that the blue dots are barely visible as they overlap with the results from the train set. The red dots represent the cosine similarity of the randomly generated key compared to the original key. The cosine similarity remains consistently low, as it is at approximately 0 for all randomly generated keys, indicating that no false positives occurred as desired by the \hyperref[req:r4:reliability]{Reliability} requirement from~\cref{sec:problemstatement}. As expected, we can see, that the H-Model exhibits poor performance, akin to that of a \naive classifier (around 10\%) due to the usage of false keys, which is represented by the green and blue dots at the bottom of ~\cref{fig1:subfigA}. The drop in accuracy from around \emph{70\%} to \emph{10\%} partly fulfills the \hyperref[req:r1:robustness]{Robustness} requirement from~\cref{sec:problemstatement}. Meanwhile, for the P-Model, a comparable outcome is yielded as depicted in \cref{fig1:subfigB}. Again, the green and blue dots represent the training set and test set accuracies, while the red dots represent the PAC (cf. ~\cref{subsec:protectionLayerInstantiations}) metric. The model exhibits poor accuracy of around \emph{30\%} (compared to \emph{70\%} for the original key), while the PAC (cf. \cref{subsec:protectionLayerInstantiations}) remains at \emph{0.5}. The PAC value can be explained by the fact that a large sample of numbers from a randomly and uniformly distributed population is taken. Therefore, the average will tend to be close to the center of the distribution. We conclude that the PAC value must be above \emph{0.5} for reasonable ownership claims. In summary, we have shown, that replacing the key with random keys yields poor model performance as well as low key similarity. Showing that both protection layers adhere to the protection property (cf.~\cref{sec:approach}).

\vspace{0.1cm}
\noindent \textbf{Incremental Key Value Replacement.} An attacker may replace a single value instead of the entire key, with the goal of decreasing key similarity while maintaining high performance. Therefore, we show that by iteratively replacing key values with random values, either the model performance degrades or the key's similarity remains high. The generated key values were again in a valid range. Thus, we performed an evaluation that involved iteratively replacing zero to 100 randomly chosen values within the original key with randomly generated values, e.g., the first iteration did not change any values and each subsequent iteration modified another random value. We limited the evaluation to 100 generated keys since the trend was already apparent. The results are depicted in \cref{fig1:subfigC}, here the red line indicates the cosine similarity compared to the reference key. The green and blue lines represent the training and test set accuracies of the H-Model. It is visible that the accuracy drops from \emph{69\%} with the original key to around \emph{56\%} with just three values replaced. For the test set an accuracy drop of the same magnitude can be observed. The key similarity remains high above \emph{0.99}. The P-Model exhibits a comparable outcome, as illustrated in \cref{fig1:subfigD}. Again, the red line represents the key's PAC, while the green and blue lines denote the train and test set accuracies. The accuracy declines from over \emph{75\%} to around \emph{27\%} for both the training and testing sets. Still, a PAC of more than \emph{0.66} is yielded. For a PAC of \emph{0.85} the model accuracy dropped to below \emph{43\%}. Both experiments indicate that the keys from the protection layers are sensitive to small changes, as they result in a significantly reduced model performance, indicating that the keys are entangled with the model parameters.

\vspace{0.1cm}
\noindent \textbf{Add Noise to Key.} Another method to tamper with the key is to add noise to the key's values. Thus, we add different levels of random noise in an iterative process to the keys of the protection layers. In particular, for Hadamard layers, we add randomly generated noise ten times in an interval of -0.2 to 0.2 to the key. For Permutation layers, 100 times a randomly selected key value is increased by one. The results for the H-model are shown in \cref{fig1:subfigE}. The meaning of the colors in the figure is identical to the experiments for the H-model. The leftmost point in ~\cref{fig1:subfigE} shows the result where no noise was added. In subsequent experiments, the added noise was further increased. In total, the random noise was added ten times to the H-Model, which prompted a decrease in accuracy by about \emph{32\%} to a value below \emph{37\%}, while the cosine similarity of the keys remained high at more than \emph{0.84}. The results for the P-Model are shown in \cref{fig1:subfigF} (the meaning of the colors is identical to previous P-Model experiments). Further incrementing random key values up to 100 times prompts a drop in accuracy from around \emph{69\%} to \emph{54\%} for the training set and \emph{54\%} to \emph{49\%} for the test set. The PAC remains at above \emph{0.98}. Both models appear to exhibit good resilience against adding noise to the key values, as an accuracy drop is clearly visible.

\vspace{0.1cm}
\noindent In summary, the experiments indicate a strong correlation between the key similarity and model performance. A dissimilarity between modified and original keys leads to a drop in model accuracy. Furthermore, the key metrics only exhibit high similarity in case very similar keys are utilized, fulfilling the \hyperref[req:r4:reliability]{Reliability} requirement from~\cref{sec:problemstatement}. We conclude that our approach is effective and reliable as the models achieve high accuracies when the correct keys are utilized. However, the model performance significantly declines and the key similarity is poor if the keys significantly deviate from reference keys, fulfilling the \hyperref[req:r1:robustness]{Robustness} requirement from~\cref{sec:problemstatement}. Based on our experiments, we suggest that a similarity value greater than \emph{0.8} for both metrics could be used as a guideline for ownership verification. This value provides a good balance between robustness and reliability. However, the threshold is an insensitive parameter.

\begin{table}
    \centering
    \begin{tabular}{cccc}
    \toprule
    Model     & Unprotected     & Hadmard & Permutation     \\
    \midrule
    CNN       & 99.826\%          & 98.795\%  & \textbf{99.861\%} \\
    \resneteighteen~\cite{resnet} & \textbf{98.894\%} & 97.788\%  & 97.756\%       \\
    \bottomrule
    \end{tabular}
    \caption{Accuracy results for CNN trained on \mnist~\cite{mnist} dataset and \resneteighteen~\cite{resnet} trained on \cifar~\cite{cifar}.}
    \label{tab:fidelityResults}
\end{table}

\subsection{Fidelity and Efficiency}
\label{sec:eval:fidelity}
In the following we first determine the best key range for the Hadamard protection layer, exhibiting the best performance and robustness. Afterward, we present evaluation results for measuring the overhead induced by the protection method.

\vspace{0.1cm}
\noindent \textbf{Key Range.}
We confirm the key parameters for the Hadamard protection layer, by evaluating the potential influence of exploding and vanishing gradients~\cite{gradientproblems} associated with high and low values (cf.~\cref{sec:approach}). Additionally, we assessed the extent to which the inclusion of positive and negative values impacted the performance of the protection layer. All experiments were conducted with a \resneteighteen~\cite{resnet} model on the \cifar dataset~\cite{cifar}.

\vspace{0.1cm}
\noindent Our experiments show that Hadamard Protection layers with high key values (generated randomly between -10.0 to 10.0) do not impact the protective capabilities. Nevertheless, they resulted in a \emph{2.57\%} decrease in accuracy compared to regular keys ranging from -1.0 to 1.0, and therefore, should be avoided. Similarly, low key values (generated randomly between -0.1 to 0.1) also resulted in a decrease of accuracy by \emph{2.11\%}. The model performance is increased by \emph{7.38\%}, using only positive values ranging from 0.0 to 1.0. However, when replacing the original key with randomly generated ones, the model protected with the key that did not include negative values (range from 0.0 to 1.0) exhibited about \emph{5\%} higher accuracy when used with random keys. Thus, the robustness, e.g., the drop in accuracy when utilized with false keys is positively impacted by including negative values. Therefore, we argue that inclusion of negative values is beneficial due to increased resilience. We conclude that adding negative values enhances the entanglement between the key and model parameters, resulting in increased robustness. Thus, all experiments are conducted with keys ranging from -1.0 to 1.0.

\vspace{0.1cm}
\noindent \textbf{Fidelity.} To measure the accuracy impact, we first trained two unprotected baseline models: A \resneteighteen~\cite{resnet} on \cifar~\cite{cifar} trained for 30 epochs and a CNN model trained on \mnist~\cite{mnist} for 30 epochs. The training duration was extended to demonstrate that the models are fully optimized, meaning a high level of accuracy has been achieved with little potential for further improvement. Next, we included Hadamard and in the second run Permutation layers into the models and again trained in the same manner. The results for the training accuracies are visualized in Appendix ~\cref{fig:baseline:hada:perm}. The figure indicates a minimal difference between the models. In the following, we report the evaluation of the whole dataset (train and test combined) reported in \cref{tab:fidelityResults}. After 30 training epochs the unprotected \resneteighteen~\cite{resnet} model achieved an accuracy of \emph{98.89\%} while the configuration with Hadamard and Permutation layers achieved accuracies of \emph{97.79\%} and \emph{97.76\%}. Accordingly, we have observed that both protection layers cause a performance drop of around \emph{1\%}, which we consider negligible as this can also stem from training randomness. Using another random seed leads the unprotected \resneteighteen~\cite{resnet} model to achieve an accuracy of \emph{97.20\%} which shows that performance fluctuations occur. Similarly, the overhead introduced by the Hadamard Protection layers in the CNN model is around \emph{1\%} and the version with Permutation layers performs \emph{0.035\%} better than the baseline. The impact of protection layers is more pronounced during the initial stages of training. However, their influence diminishes as training progresses, and ultimately, models with and without protection layers achieve almost identical accuracies. Both models, in all three configurations, display a similar training accuracy curve and accuracy values, leading us to believe that the impact of our approach on the model performance is negligible fulfilling the \hyperref[req:r3:efficiency]{Efficiency} requirement from~\cref{sec:problemstatement}.

\subsection{Generalizability}
To demonstrate the generalizability, we protect a \resneteighteen~\cite{resnet} model using Hadamard layers (H-ResNet) followed by another version utilizing Permutation layers (P-ResNet). We trained both model variations on the \cifar dataset~\cite{cifar} and the \gtsrb dataset~\cite{gtsrb}. The H-ResNet achieved a training set accuracy of \emph{80.48\%} on \cifar~\cite{cifar} and an accuracy of \emph{94.89\%} on \gtsrb~\cite{gtsrb}. Furthermore, the P-ResNet achieved an accuracy of \emph{82.06\%} on \cifar~\cite{cifar} and \emph{97.18\%} on \gtsrb~\cite{gtsrb}. To ascertain the robustness for the H-ResNet and P-ResNet trained on \cifar~\cite{cifar}, we again conducted the same experiments as in \cref{eval:subsec:generalfunc}. In particular, three experiments were conducted. The first experiment replaced the original key with 100 randomly generated keys. The second experiment replaced a single key value, and the third experiment added noise to the key. The results are depicted in Appendix \cref{fig:subfigures:ResNetFunctionality} and show that the \resneteighteen~\cite{resnet} model behaves very similar to the previously assessed CNN model. Therefore, we conclude that both protection layers are applicable to multiple different datasets and model architectures.

\begin{table}
\centering
\resizebox{0.5\textwidth}{!}{

  \begin{tabular}{lccccc}
  \toprule
    & & \multicolumn{2}{c}{All Params} & \multicolumn{2}{c}{Key Params} \\
    & & Same LR & \nicefrac{1}{10} of LR & Same LR & \nicefrac{1}{10} of LR \\
    \midrule
    (1) & \textbf{COS} & \textbf{0.9975} & \textbf{0.9999} & \textbf{0.9855} & \textbf{0.9994} \\
    (2) & Test Set Acc & 97.14\% & 93.28\% & 76.03\% & 64.63\% \\
    (3) & $\Delta$ Train Acc & -13.27\% & -1.74\% & -2.63\% & -1.64\% \\
    \bottomrule
\end{tabular}
}
\caption{Fine-Tuning accuracy (Acc) for \resneteighteen~\cite{resnet} protected using Hadamard layers. The table depicts results for all parameters and only key parameters being tuned, both with the same learning rate (LR) and \nicefrac{1}{10} of the LR.}
\label{tab:finetuning:hresnet}
\end{table}

\vspace{0.1cm}
\noindent \textbf{Applicability to Transformer-Based models}
To demonstrate the applicability of \sysname~to Transformer-based models, e.g. Large Language Models~\cite{attention, llama, gpt3}, we protect a Vision Transformer~\cite{visiontransformer}. Transformer models are naturally divided into processing blocks. Our model consists of 7 Transformer blocks. The embedding size and the Fully-Connected (FC) layers input dimension of the Transformer blocks are 384. Further, the model utilizes 12 Self-Attention modules. We protect the model by inserting a Hadamard Protection layer after the first FC Layer within each Transformer block. The unprotected Vision Transformer achieved an accuracy of 83.45\% with protection layers in place. Similar to the \resneteighteen~\cite{resnet} experiments, we evaluated the robustness of \sysname~with three experiments. The first experiment was repeated 100 times and evaluated the model's performance using randomly generated keys, mimicking an adversary trying to utilize the model with a different key. Each time, a significant performance drop of around 70\% occurred. Incrementally replacing single key values (up to 100) resulted in a negligible performance drop (less than 1\%), however, the cosine similarity of the key remained above 99.94\%. Finally, adding noise to the key until the cosine similarity dropped to 80\% caused a performance drop to around 20.59\% (a decrease of approximately 62\%). These results are very similar to the other models exhibiting the same behaviour. Thus, \sysname~can be effectively applied to Transformer-based architectures as well.

\subsection{Robustness and Reliability}
\label{subsec:eval:robustness}
In this section we evaluate the robustness of our approach against more sophisticated adversarial modifications. Common attacks, such as Fine-Tuning (cf.~\cref{subsec:background:dnnipprotectionattacks}) and Pruning (cf.~\cref{subsec:background:dnnipprotectionattacks}), can only be applied for the Hadamard Protection layer since the Permutation Protection layer cannot be optimized using gradient descent techniques. This is due to the Permutation Protection layer simply rearranging the order of outputs without modifying their values.

\vspace{0.1cm}
\noindent However, the Hadamard layer introduces key values (parameters) that adversaries can directly optimize using gradient descent, making it applicable to common attacks such as Pruning and Fine-Tuning. Therefore, we evaluated the vulnerability of the Hadamard Protection layer to these two scenarios.

\vspace{0.1cm}
\noindent \textbf{Fine-Tuning.} In fine-tuning attacks (cf.~\cref{sec:problemstatement}), attackers alter the parameters of trained models to circumvent security measures while preserving the accuracy. \sysname~is not vulnerable to such attacks because its security is dependent on the key layers that remain unchanged even if other model parameters are manipulated. We considered two scenarios how malicious actors may adapt Fine-Tuning to \sysname. In the first, adversaries treat the values in the key layer as if they belong to the model parameters, e.g., make them trainable. The goal is to compromise the protection mechanism by Fine-Tuning all model parameters and key values. Second, adversaries freeze all model parameters except for the key values, which are then fine-tuned in order to alter them while preserving the accuracy. The third scenario, training only the model parameters without the key values, is not beneficial, as in this case, the model identifier would remain unchanged.

\vspace{0.1cm}
\noindent To evaluate \sysname's robustness against adapted Fine-Tuning scenarios, we continued training the H-ResNet on the \cifar~\cite{cifar} test set. We trained for another ten epochs with the same learning rate, as well as with \nicefrac{1}{10} of the original learning rate (similarly to~\cite{uchida2017embedding,chen2019deepmarks,xie2021deepmark,lemerrer2019frontierstitching,adi2018turning,li2021spreadtransform}). The evaluation results are depicted in \cref{tab:finetuning:hresnet}. In all tested scenarios the cosine similarity between the modified key and the original key was above \emph{0.98} as can be seen in line 1 of \cref{tab:finetuning:hresnet}. Furthermore, line 2 reports the accuracies of the test set which was used for Fine-Tuning. Line 3 of ~\cref{tab:finetuning:hresnet} shows the change in the training set accuracy. Therefore, we argue that the approach is robust against Fine-Tuning a protected model. To further evaluate the robustness of our approach we fine-tune a \resneteighteen~\cite{resnet} model, protected using Hadamard layers, and trained on \gtsrb~\cite{gtsrb} for 10 epochs on the \cifar dataset~\cite{cifar} for another 10 epochs. This process involved replacing the last layer of the model as the number of output classes changed, therefore, we only evaluated the scenario where all model parameters including those of the key layer are fine-tuned. The learning rate was set to \nicefrac{1}{10} of the original learning rate as for higher learning rates the accuracy results degraded. After the Fine-Tuning, the cosine similarity of the key was above \emph{0.99} while the train and test accuracies of the \gtsrb~\cite{gtsrb} dataset were \emph{86.78\%} and \emph{94.18\%} respectively. Our results show that an adversary can not remove the identifier by adapting the model to his purposes. The approach exhibits strong resilience against adaptive adversaries performing Fine-Tuning attacks, essentially fulfilling the \hyperref[req:r1:robustness]{Robustness} requirement from~\cref{sec:problemstatement}.

\begin{figure}
        \captionsetup[subfigure]{aboveskip=-2pt,belowskip=-2pt}
	\begin{centering}
        \quad\quad
        \begin{subfigure}{0.35\linewidth}
            \includegraphics[width=\linewidth]{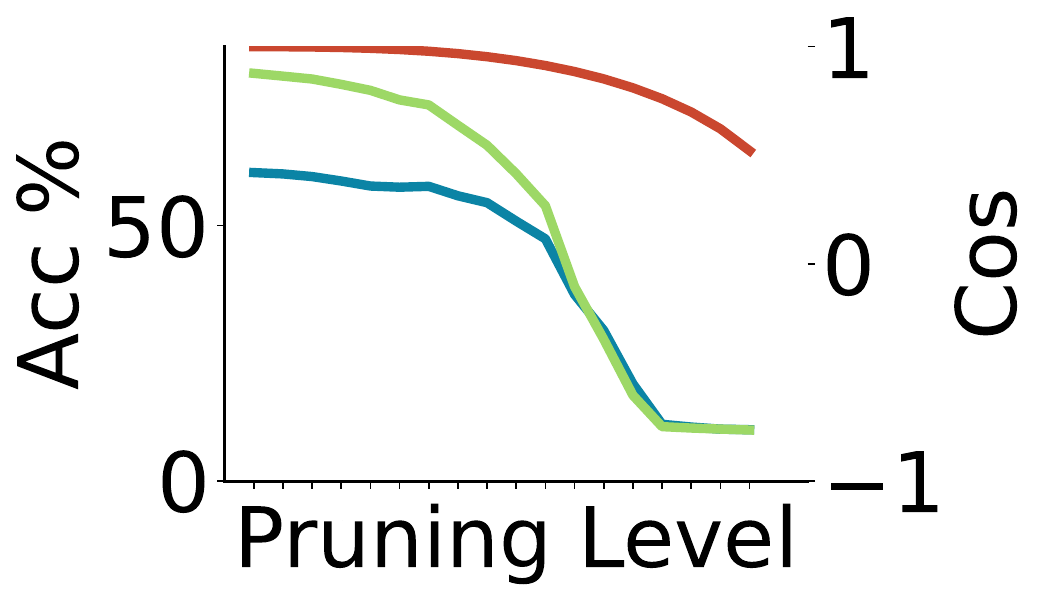}
            \caption{All Parameters}
            \label{fig:subfig4:a}
	\end{subfigure}
        \hfill
	\begin{subfigure}{0.35\linewidth}
		\includegraphics[width=\linewidth]{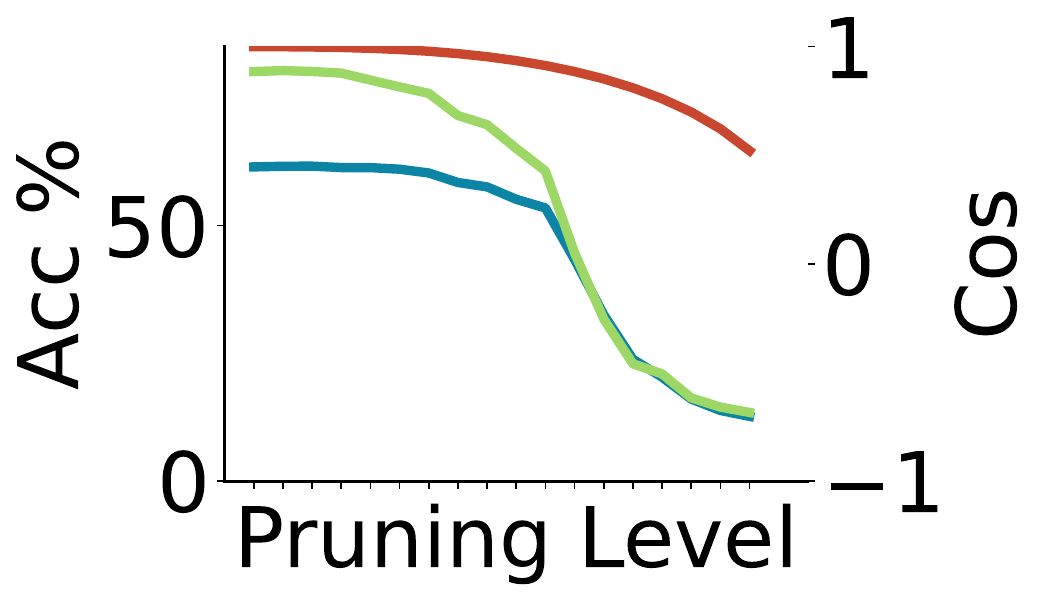}
		\caption{Key Parameters}
		\label{fig:subfig4:b}
	\end{subfigure}
         \quad\quad
	\end{centering}
        \caption{Pruning from 5\% to 90\% for the H-ResNet trained on \cifar~\cite{cifar}. The Red line depicts the cosine similarity, while the Green/Blue lines depict train/test accuracy.}
	\label{fig4:pruning}
\end{figure}

\vspace{0.1cm}
\noindent
\textbf{Pruning.}
Besides Fine-Tuning, we investigate another common attack called Pruning (cf. \cref{sec:problemstatement}) similar to~\cite{uchida2017embedding,rouhani2019deepsigns,chen2019deepmarks,xie2021deepmark,lemerrer2019frontierstitching,li2021spreadtransform}. We again employ the two scenarios from Fine-Tuning, as Pruning only the model parameters would leave the key layers values untouched. The results are depicted in \cref{fig4:pruning}. The blue line depicts the test set accuracy, while the green line represents the training set accuracy. The red line depicts the cosine similarity of the key. First, we prune all model parameters including those of the key layers and second, we only prune the values from the key layers. We used the \resneteighteen~\cite{resnet} model which was trained on \cifar~\cite{cifar} for ten epochs and progressively pruned the model parameters and the key layers. We started with Pruning \emph{5\%} of the lowest values and increased the Pruning value by \emph{5\%} until we reached \emph{90\%}. As shown in \cref{fig:subfig4:a}, in case all parameters are pruned, for a key similarity of \emph{0.85} (reached at a Pruning level of \emph{65\%}) the accuracies are at \emph{27.76\%} for the test set and \emph{29.5\%} for the train set. Similarly, as shown in \cref{fig:subfig4:b}, in case only the key parameters are pruned for a key similarity of \emph{0.85} (also reached at a Pruning level of \emph{65\%}) the accuracies are at \emph{31.73\%} for the train set and \emph{32.57\%} for the test set. We conclude, that in both scenarios the model performance is correlated with the similarity of the key. Hence, the approach is resilient against Pruning attacks, adhering to the \hyperref[req:r1:robustness]{Robustness} requirement from~\cref{sec:problemstatement}.

\vspace{0.1cm}
\noindent
\textbf{Adaptive Adversary.}
An adaptive adversary could integrate the protection layer's parameters into the preceding FC layer (see FCN model in \cref{app:modeloverview}). This attack strategy was discussed in \cref{sec:security_analysis} and illustrated in~\cref{fig:adaptivemerge}, where we determined that the key could be extracted from the merged parameters using the process outlined in \cref{app:keyextraction}. A FCN model was trained on the \mnist dataset~\cite{mnist}. Next, both Hadamard Protection layers from the FCN model were multiplied into the parameters of the preceding FC layers. Subsequently, the model underwent Fine-Tuning on the test dataset using the same learning rate used in the regular training phase. Our evaluation results show that even after ten Fine-Tuning iterations, the key's values could be extracted from the model parameters with a cosine similarity exceeding \emph{0.99}. Therefore, we conclude that multiplying the Hadamard key values into parameters of a FC layer is not an effective means of circumventing the protection mechanism. To ensure that unprotected model do not yield high key similarity in case the previously described extraction process is performed, we conducted a second experiment regarding retrieval of the key from an unprotected and trained model. Here, the cosine similarity was very low at \emph{-0.13\%}, therefore, we consider the \hyperref[req:r4:reliability]{Reliability} requirement from~\cref{sec:problemstatement} fulfilled.

\begin{figure}
    \centering
    \includegraphics[width=0.35\linewidth]{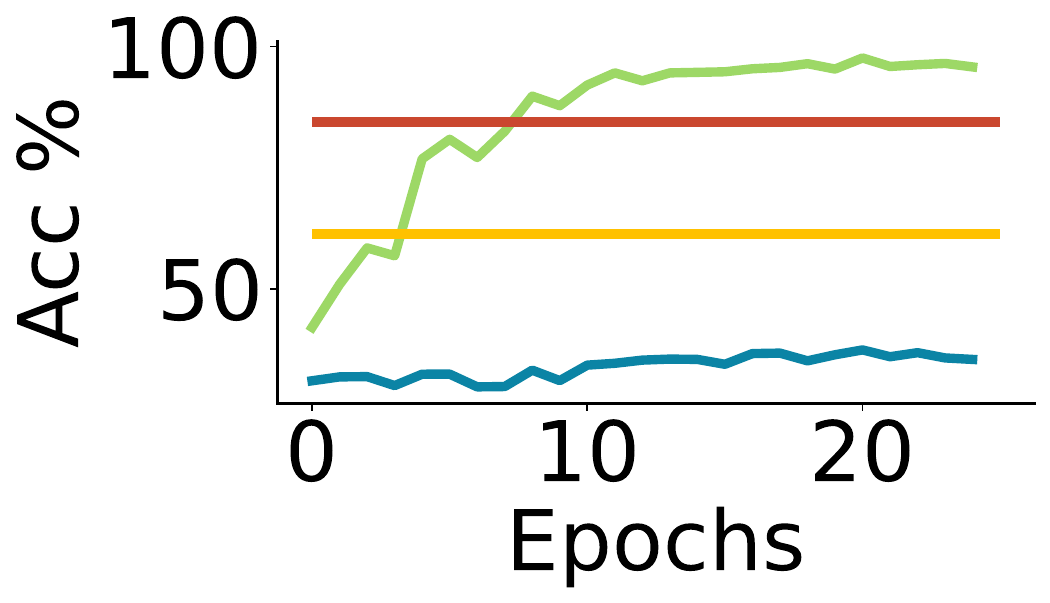}
    \caption{Model test accuracy (green line) where key layers were substituted with FC layers. The blue line depicts the training accuracy, the Red/Yellow line indicates the baseline performance from the protected model on the train/test set.}
    \label{fig:eval:linearsubstitute}
\end{figure}

\vspace{0.1cm}
\noindent To further assess the effectiveness of replacing protection layers with FC layers, we conducted an additional experiment. While replacing Permutation layers is ineffective, as ideally, they would simply learn how to rearrange the output values in a manner already accomplished by the Permutation layer, replacing Hadamard protection layers with FC layers holds limited promise. The idea was introduced in~\cite{PassportingAttack}. It is noteworthy that replacing each protection layer in the \resneteighteen~\cite{resnet} model with a FC layer introduces a parameter overhead of 28 times, e.g., the amount of parameters is increased from 11M to 318M. The overhead renders the attack impractical in real-world scenarios. Nevertheless, we pursued this experiment to evaluate whether FC layers could effectively learn the scaling introduced by Hadamard layers. This process entailed Fine-Tuning a \resneteighteen~\cite{resnet} model on the test set using the identical learning rate used in model training. As illustrated in \cref{fig:eval:linearsubstitute}, the substitution of protection layers with FC layers attains peak performance on the test set depicted by the green line, surpassing the baseline test accuracy depicted in yellow. However, the attack is unable to achieve performances close to the baseline accuracy of \emph{80.48\%} (depicted as the red line) for the training set, represented by the blue line. Even after 25 epochs, the accuracies of the training set persist at low levels of approximately \emph{35\%}, corresponding to a decline of roughly \emph{45\%}. This indicates that the model is over-fitting to the test set. We conclude that replacing the protection layers with FC layers is not a viable option to circumvent the protection.

\subsection{\sysname's Runtime}
To measure the runtime overhead of \sysname~, we trained three models for 10 epochs while averaging their time per epoch. We first trained the CNN model on the \cifar~\cite{cifar} dataset without protection layers. The mean time (seconds) per epoch was \emph{3.1384s} with a variance of \emph{0.031}. The model protected using two Hadamard layers had a mean time per epoch of \emph{3.2745s} with a variance of \emph{0.019}. The model protected using two Permutation layers had a mean time per epoch of \emph{3.5352s} with a variance of \emph{0.052}. These results demonstrate that the overhead introduced by \sysname~is minimal, e.g., less than \emph{5\%} for the model protected using Hadamard layers.

\section{Discussion}
\label{sec:discussion}
The following section presents a discussion on the placement, combination, and amount of protection layers. Additionally, we will discuss the publication of the key.

\vspace{0.1cm}
\noindent \textbf{Protection Layer Placement.} \noindent The placement of protection layers within a DNN can impact its robustness against adversarial attacks. Placing them after activation functions can make it more difficult for adversaries to merge them into the preceding unprotected layer. This is because activation functions introduce non-linearity, making it more challenging to integrate them into protection layers.

\vspace{0.1cm}
\noindent \textbf{Protection Layer Combination.} The Hadamard and Permutation layers (cf. ~\cref{subsec:protectionLayerInstantiations}) are two distinct instantiations that offer unique advantages. The Hadamard layer introduces a scaling pattern. The permutation layer, on the other hand, changes the order of output values. Combining both types of protection layers within a single DNN might further complicate the task of removing the protection layers for an adversary.

\vspace{0.1cm}
\noindent \textbf{Protection Layer Amount.} The number of protection layers used in a DNN has a direct resilience impact. Adding more protection layers can improve the robustness of the protection methods and increase the confidence in the model identifier. However, too many protection layers can lead to drawbacks, such as longer runtime. Therefore, achieving the desired level of robustness and identification confidence requires a balanced number of protection layers.

\vspace{0.1cm}
\noindent \textbf{Key Publication.} To facilitate efficient ownership verification, the model owner must establish that the key integrated into the model was indeed chosen by them and not another party. Therefore, the key or the hash of it should be publicly disclosed prior to model training, such as on a platform that assigns verifiable timestamps and links the key to the owner of the model, e.g. a blockchain~\cite{timestampingonblockchain}.

\section{Related Work}
\label{sec:relatedwork}
This section presents DNN IP protection works related to \sysname.

\vspace{0.1cm}
\noindent \textbf{Watermarking.} Watermarking (WM) (cf.~\cref{sec:background}) is a related approach to IP protection for DNNs. White-box WM schemes~\cite{uchida2017embedding, li2021spreadtransform, tartaglione2021delving, wang2020watermarkingbackpropagation, wang2021riga, rouhani2019deepsigns}, which directly modify and analyze the model weights, are among the most prevalent DNN WM techniques. These approaches typically necessitate a secret key and a secret watermark that can be extracted using this key. Consequently, the approaches can only be employed once, as the secret key must be revealed for ownership verification. Furthermore, watermark embedding is primarily accomplished through the use of an additional loss function, which increases training complexity. In contrast, our approach does not rely on any secrets, enabling limitless ownership verification. Additionally, the approach does not significantly increase training overhead as it does not introduce additional trainable parameters.

\vspace{0.1cm}
\noindent The second category, Black-box WM schemes~\cite{lemerrer2019frontierstitching, adi2018turning, zhang2018protecting, zheng2019blindwatermark, guo2018watermarking, jia2021entangledwm}, involve feeding specific input samples into the model and analyzing their outputs. These inputs, function as the secret keys and are kept confidential. As a result, often the same limitations apply to these approaches. To incorporate the watermark, the training data must be manipulated which could degrade the models performance.

\vspace{0.1cm}
\noindent \textbf{Fingerprinting.} Fingerprinting (FP) extracts unique identifiers from trained DNNs for ownership verification. The methods can be categorized as parameter-based or input-based. Both have similar limitations than WM schemes. Parameter-based FP~\cite{rouhani2019deepsigns,chen2019deepmarks} relies on a secret key, limiting its applicability. In contrast, \sysname~does not rely on secrecy, enabling an arbitrary number of verifications.

\vspace{0.1cm}
\noindent Input-based FP~\cite{yang2022metafinger, lukas2021dnnfingerprinting, xiaoyu2021ipguard} utilizes distinct output patterns generated by different DNNs for specific inputs to identify a model. However, this approach requires generating carefully crafted inputs, which can be used for adaptive attacks and, thus, need to be kept secret. Additionally, the identifier is only extracted after model training, leaving intermediate model states, known as checkpoints, susceptible to misuse by adversaries. In contrast, \sysname~provides a static identifier, unchanged during model training, addressing these limitations. DeepJudge~\cite{chen2022deepjudge}, a testing framework proposed as alternative to traditional WM techniques, faces similar limitations. It compares the behavioral similarities between a trained DNN and a potentially infringing model based on six metrics. These metrics are derived from selected inference samples that capture the models' characteristics. However, the output for these specific samples could be manipulated and, contrary to \sysname, the method cannot safeguard intermediate model states, leaving them susceptible to misuse.

\vspace{0.1cm}
\noindent \textbf{Passporting.} 
Fan \etal~\cite{PassportingRevisedVersion,PassportingInitial} introduces passport layers as a DNN model protection mechanism. However, this approach is limited to specific convolutional layer configurations, namely pairs of convolutional layers followed by a normalization layer. The normalization layers are modified to form passport layers, functioning similar to our Hadamard layers. These layers incorporate trainable parameters that are optimized alongside the remaining model parameters and constrained by an additional loss. Therefore, the training complexity is increased and the application of the method results in training overhead. Furthermore, it prevents the protection of intermediate model checkpoints, as the passport layer parameters change throughout training. In contrast, \sysname~can be applied to both convolutional and Fully-Connected layers, regardless of subsequent layers, without introducing additional trainable parameters. Moreover, it safeguards intermediate checkpoints. Zhang \etal~\cite{passportawarenorm} introduce a method that incorporates a secret passport layer during training alongside the unprotected model. After releasing the unprotected model, the secret passport layers can be employed analogously to a secret key to verify ownership. This is because the unmodified model, when equipped with the secret passport layer, exhibits unique behavior. A similar approach is also proposed as an alternative in DeepIPR~\cite{PassportingRevisedVersion}. In contrast, our approach does not rely on any secret keys. Additionally, Chen \etal~\cite{PassportingAttack} demonstrated that DeepIPR's passport layers are vulnerable to ambiguity attacks. These attacks attempt to falsely claim ownership by replacing the passport's trainable parameters with different ones. A small portion (10\%) of training data is leveraged with additional fully connected layers to identify alternative parameters. We have shown that finding alternative keys that have significant dissimilarity is infeasible for our approach. Furthermore, an adversary could attempt to circumvent the protection mechanism by replacing the protection layers with fully connected layers. However, our experiments in~\cref{subsec:eval:robustness} have shown that this strategy is ineffective against \sysname~and does not scale well in terms of model size.

\vspace{0.1cm}
\noindent \textbf{Hardware-based IP protection.}
Hardware-based IP protection methods can be broadly classified into two categories: Hardware-Level IP protection~\cite{chen2019deepattest} employs trusted execution environments (TEEs) to ensure that only approved models can be executed on specific hardware devices. A TEE acts as a secure enclave within the hardware, safeguarding sensitive information. Only models that match the secret model identifier are granted access to the hardware. Additionally, Hardware-Assisted IP protection~\cite{HardwareAssistedIP} involves encrypting portions of the model and executing them within a TEE. During inference in the TEE a secret key is employed. In contrast, \sysname~eliminates the need for dedicated hardware or secret keys.

\section{Conclusion}
\label{sec:con}
Protecting the IP rights of DNN creators is a significant challenge in the rapidly evolving field of ML. Current approaches modify the training dataset, rely on the secrecy of an embedded key, introduce additional training parameters, restrict themselves to specific layer types, or leave unfinished model checkpoints unprotected. To address these limitations, we introduce \sysnamelongw, a novel protection method that integrates protection layers into the model's architecture. These layers allow for accurate identification of the model, enabling ownership claims by the creator. Our approach includes two instances of protection layers, both demonstrating high resilience against fine-tuning, pruning, and sophisticated adaptive adversarial attacks while incurring negligible performance and runtime overhead. We extensively evaluated the approach across three datasets and three model architectures, confirming the efficacy of \sysnamelong~in safeguarding DNNs.
\clearpage

		
		
		%
		\bibliographystyle{IEEEtranS}
		\bibliography{references}
		
		\appendices
		\section{Hadamard Key Merge into Linear Layer}
\label{app:weightmerge}
In case no activation function is utilized after calculation of the linear layers output it is straightforwardly possible to merge the key of the Hadamard layer into the weights of the linear layer. Assume an input vector $x$, a linear layer with weights $w$ and bias $b$, followed by a Hadamard protection layer with key $k$. The forward pass is computed as $y = (x \cdot w^{t} + b) \times k$ which expands to $(\sum_{i}{x_{i,j} \cdot w_{i,j}} + b_j) \times k_j$, where $i$ denotes the i-th and $j$ denotes the j-th column in the matrix. Now an adversary can compute $w_{i,j}' = w_{i,j} \times k_j$ and $b_j' = b_j \times k_j$ to merge the key with the linear layer. The weights and biases of the new linear layer are now specified by $w'$ and $b'$.

\section{Hadamard Key Extraction from Merged Linear Layer}
\label{app:keyextraction}
The key can be extracted from a merged linear layer by calculating $\frac{w'_{i,j}}{w_{i,j}}=k_{i,j}$. For each key value multiple values are obtained, therefore,  an averaging mechanism is required to obtain a single value. Also the key from the bias is extracted by calculating $\frac{b'_{j}}{b_{j}} = k_j$. To enhance the robustness and perform aggregation, outliers of three times the standard deviation from the mean are removed and then the mean of the remaining values is computed. The key from the weights and bias are equally weighted and averaged. Therefore, the resulting key can be compared as described in \cref{sec:approach}.

\section{Hadamard Key Merge into Convolutional Layer}
\label{app:cnnmerge}
Consider a convolutional layer with one filter with weights $f$, a kernel size of 2 by 2 and for simplicity without bias, after calculation of the convolution layer a Hadamard layer with key $k$ is applied. The forward pass calculates $y  = (x * f)$, where $*$ denotes the cross-correlation operation and $x$ is the input vector as shown in \cref{xinputmatrix}. The value from the convolutions output are calculated as $x_1 * f_1 + x_2 * f_2 + x_4 * f_3 + x_5 * f_4$, the second cell is calculated as $x_2 * f_1 + x_3 * f_2 + x_5 * f_3 + x_6 * f_4$. Now to merge the key layer into the first convolutional filters weight one would need to calculate $f_1' = f_1 \times k_1$ and $f_1' = f_1 \times k_2$. As $k_1 \neq k_2$ it is not possible to merge the key layer into the convolutional filters weights. This assumes that the values of the keys are not equivalent, which is highly probable given that the values are randomly generated.

\begin{figure}[ht]
\begin{center}
\begin{math}
\begin{bmatrix}
x_1 & x_2 & x_3\\
x_4 & x_5 & x_6
\end{bmatrix}
\end{math}
\end{center}
\caption{Input Matrix $x$.}
\label{xinputmatrix}
\end{figure}

\section{Model Architectures}
\label{app:modeloverview}
The approaches were assessed across different model architectures, encompassing Fully Connected and Convolutional models, alongside the widely used \resneteighteen~\cite{resnet} architecture.

\vspace{0.1cm}
\noindent\textbf{Fully Connected} The Fully Connected Network (FCN) encompasses linear layers and the ReLu activation function as shown in ~\cref{tab:app:fcn}. The model can only be protected using Hadamard protection layers.

\begin{table}[ht]
\begin{tabular}{lcc}
\toprule
Layer              & Input Width and Height & Input Channels \\
\midrule
Linear 1           & 32x32            & 3               \\
\textbf{Protection Layer 1} & 1              & 15               \\
ReLu & 1              & 15               \\ \hdashline 
Linear 2           & 1              & 15               \\
\textbf{Protection Layer 2} & 1              & 10               \\
ReLu & 1              & 10               \\ \hdashline 
Linear 3       & 1              & 10                \\
\bottomrule
\end{tabular}
\caption{Architecture of the Fully Connected Network. The dimensions are calculated based on usage with the \cifar~\cite{cifar} dataset.}
\label{tab:app:fcn}
\end{table}

\vspace{0.1cm}
\noindent\textbf{Convolutional}
The architecture of the convolutional model is shown in ~\cref{tab:app:cnn}. The convolutional layers (Conv2d) have kernel sizes of 5 by 5 and the maxpooling layers (MaxPool2d) have kernels of size 2 by 2. In case the model is protected by Permutation layers only the Protection Layer 1 and Protection Layer 2 are utilized, as the Permutation layers can not be used in conjunction with Linear layers. 

\begin{table}[ht]
\begin{tabular}{lccc}
\toprule
Layer              & Input Width and Height & Input Channels \\
\midrule
Conv2d             & 32x32              & 3             \\
\textbf{Protection Layer 1} & 28x28             & 32              \\
ReLu               & 28x28             & 32             \\
MaxPool2d          & 28x28             & 32              \\ \hdashline 
Conv2d             & 14x14             & 32              \\
\textbf{Protection Layer 2} & 10x10             & 64              \\
ReLu               & 10x10             & 64              \\
MaxPool2d          & 10x10             & 64              \\ \hdashline 
Linear             & 5x5          & 64             \\
\textbf{Protection Layer 3} & 1            & 512            \\
ReLu               & 1            & 512            \\ \hdashline 
Linear             & 1            & 512             \\
\textbf{Protection Layer 4} & 1            & 256             \\
ReLu               & 1            & 256             \\ \hdashline 
Linear             & 1            & 256               \\
\bottomrule
\end{tabular}
\caption{Architecture of the Convolutional Neural Network (CNN). The dimensions are calculated based on usage with the \cifar~\cite{cifar} dataset.}
\label{tab:app:cnn}
\end{table}

\vspace{0.1cm}
\noindent\textbf{ResNet-18}
The architecture of the protected \resneteighteen~\cite{resnet} model is shown in \cref{app:tab:resnet}. The model consists of convolutional layers (Conv2d), batch normalization layers (BatchNorm), and skip connections (SkipConnection) which add the input of previous layers to following layers. We protect the \resneteighteen~\cite{resnet} model with nine well-distributed Protection layers as shown in \cref{app:tab:resnet}. For Permutation Layer Protection only the first seven Protection Layers are used because protection layers 8 and 9 consists of a single value per channel which can not be permuted. The Hadamard protection layers' key values represent \emph{0.28\%} of the total parameters, while the Permutation protection layers' key values represent only \emph{0.008\%} of the total parameters.

\begin{table}[ht]
\resizebox{0.31\textwidth}{!}{
\begin{tabular}{lccc}
\toprule
Layer              & Input Width and Height & Input Channels \\
\midrule
Conv2d             & 32x32             & 3             \\
\textbf{Protection Layer 1} & 16x16             & 64              \\
BatchNorm               & 16x16             & 64             \\
ReLu          & 16x16             & 64              \\  
MaxPool2d             & 16x16             & 64              \\ \hdashline 
Conv2d               & 8x8             & 64              \\
\textbf{Protection Layer 2} & 8x8             & 64              \\
BatchNorm               & 8x8             & 64              \\
ReLu               & 8x8             & 64              \\
Conv2d               & 8x8             & 64              \\
BatchNorm               & 8x8             & 64              \\
ReLu                    & 8x8            & 64        \\  \hdashline 
Conv2d               & 8x8             & 64              \\
\textbf{Protection Layer 3} & 8x8             & 64              \\
BatchNorm               & 8x8             & 64              \\
ReLu               & 8x8             & 64              \\  
Conv2d               & 8x8             & 64              \\
BatchNorm               & 8x8             & 64              \\
ReLu               & 8x8             & 64              \\ \hdashline 
Conv2d               & 8x8             & 64              \\
\textbf{Protection Layer 4} & 4x4             & 128              \\
BatchNorm               & 4x4             & 128              \\
ReLu               & 4x4             & 128              \\ 
Conv2d               & 4x4             & 128              \\
BatchNorm              & 4x4             & 128              \\
SkipConnection          & 4x4            & 128 \\
ReLu               & 4x4             & 128              \\  \hdashline 
Conv2d               & 4x4             & 128              \\
\textbf{Protection Layer 5} & 4x4             & 128              \\
BatchNorm               & 4x4             & 128              \\
ReLu               & 4x4             & 128              \\ 
Conv2d               & 4x4             & 128              \\
BatchNorm               & 4x4             & 128              \\
ReLu               & 4x4             & 128              \\  \hdashline 
Conv2d               & 4x4             & 128              \\
\textbf{Protection Layer 6} & 2x2             & 128              \\
BatchNorm               & 2x2             & 256              \\
ReLu               & 2x2             & 256              \\
Conv2d               & 2x2             & 256              \\
BatchNorm               & 2x2             & 256              \\
SkipConnection          & 2x2 & 256 \\
ReLu               & 2x2             & 256              \\  \hdashline 
Conv2d               & 2x2             & 256              \\
\textbf{Protection Layer 7} & 2x2             & 256              \\
BatchNorm               & 2x2             & 256              \\
ReLu               & 2x2             & 256              \\ 
Conv2d               & 2x2             & 256              \\
BatchNorm               & 2x2             & 256              \\
ReLu               & 2x2             & 256              \\  \hdashline 
Conv2d               & 2x2             & 256              \\
\textbf{Protection Layer 8} & 1x1             & 256              \\
BatchNorm               & 1x1             & 512              \\
ReLu               & 1x1             & 512              \\ 
Conv2d               & 1x1             & 512              \\
BatchNorm               & 1x1             & 512              \\
SkipConnection & 1x1 & 512 \\
ReLu               & 1x1             & 512              \\  \hdashline 
Conv2d               & 1x1             & 512              \\
\textbf{Protection Layer 9} & 1x1             & 512              \\
BatchNorm               & 1x1             & 512              \\
ReLu               & 1x1             & 512              \\ 
Conv2d               & 1x1             & 512              \\
BatchNorm               & 1x1             & 512              \\
ReLu               & 1x1             & 512              \\  \hdashline 
AdaptiveAvgPool2d            & 1x1            & 512               \\
Linear            & 1            & 512               \\
\bottomrule
\end{tabular}}
\caption{Architecture of the \resneteighteen~\cite{resnet} model with Protection layers after first Convolution. The dimensions are calculated based on the \cifar~\cite{cifar} dataset.}
\label{app:tab:resnet}
\end{table}

\section{Additional Functionality Experiments}
\begin{figure}
        \captionsetup[subfigure]{aboveskip=-2pt,belowskip=-2pt}
	\centering
	\begin{subfigure}{0.32\linewidth}
		\includegraphics[width=\linewidth]{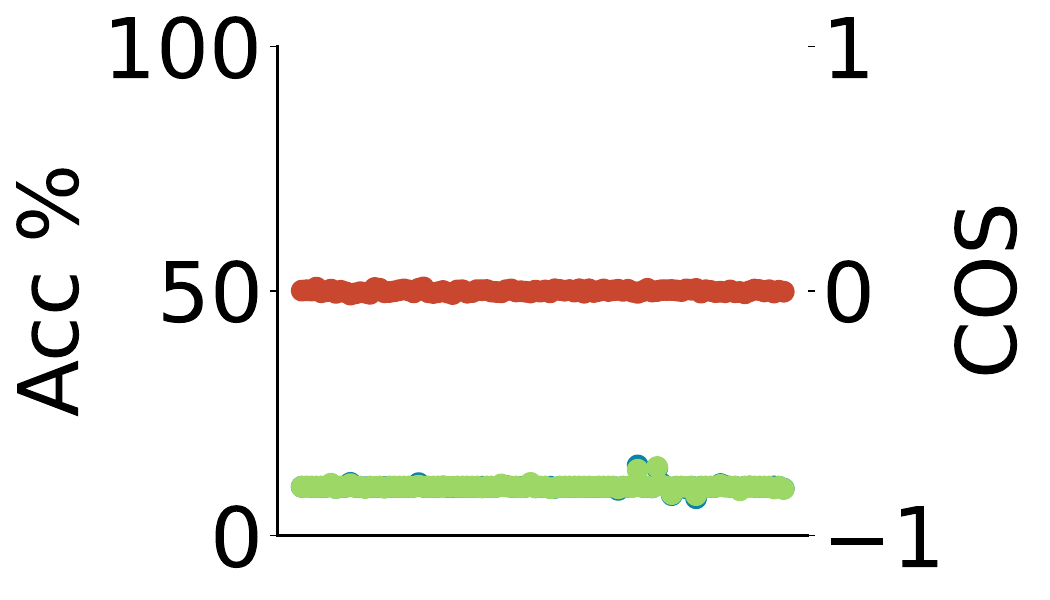}
		\caption{Random keys}
		\label{fig1:subfigRA}
	\end{subfigure}
	\begin{subfigure}{0.32\linewidth}
		\includegraphics[width=\linewidth]{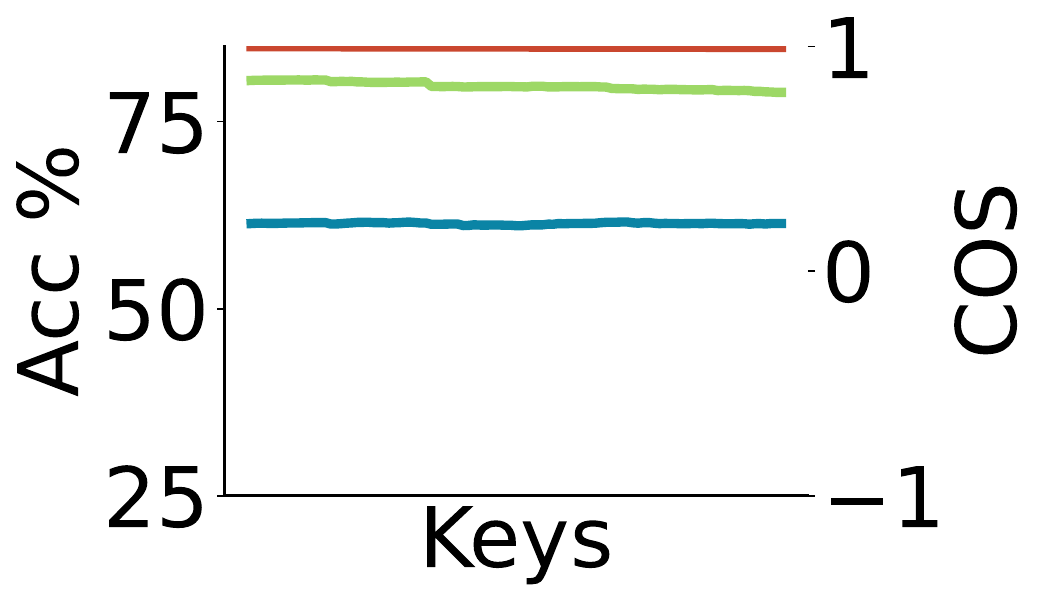}
		\caption{Random value}
		\label{fig1:subfigRC}
	\end{subfigure}
	\begin{subfigure}{0.32\linewidth}
	        \includegraphics[width=\linewidth]{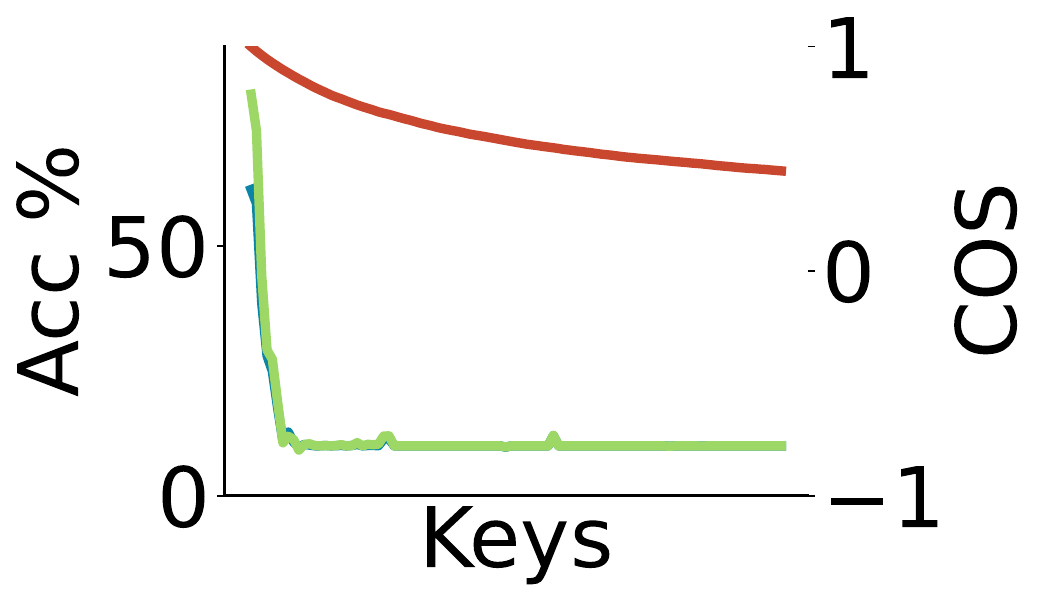}
	        \caption{Add noise}
	        \label{fig1:subfigRE}
         \end{subfigure}
         \hfill
        \vspace{0.1cm}
	\vfill
	\begin{subfigure}{0.32\linewidth}
		\includegraphics[width=\linewidth]{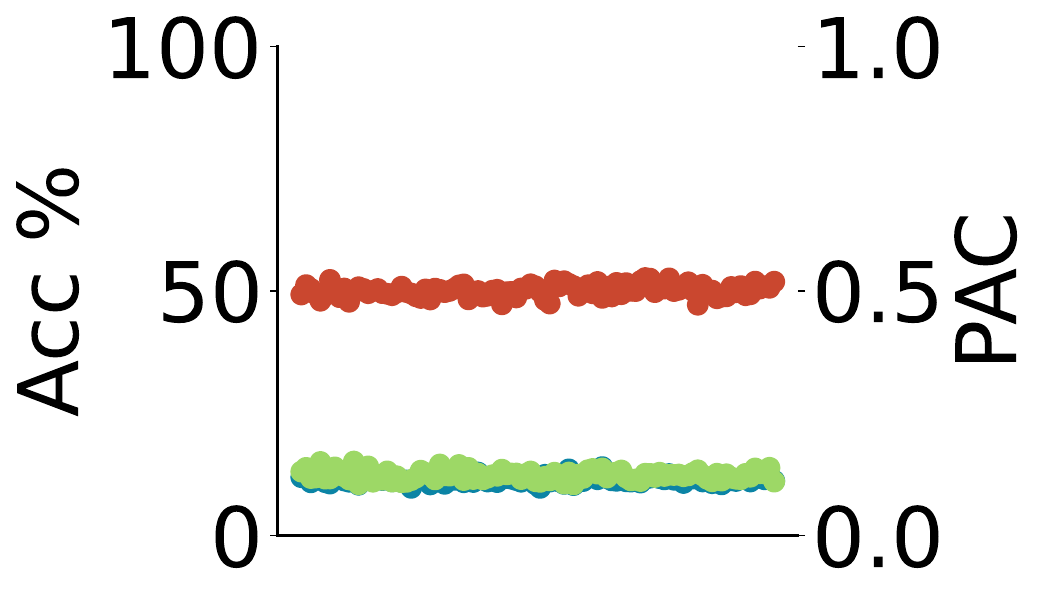}
		\caption{Random keys}
		\label{fig1:subfigRB}
	\end{subfigure}
	\begin{subfigure}{0.32\linewidth}
		\includegraphics[width=\linewidth]{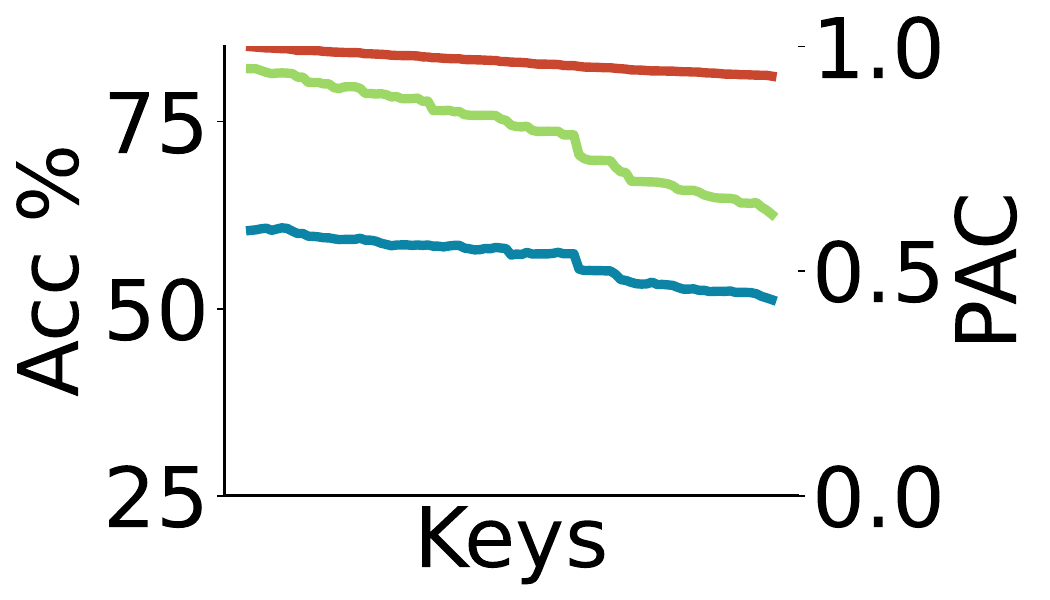}
		\caption{Random value}
		\label{fig1:subfigRD}
	\end{subfigure}
	\begin{subfigure}{0.32\linewidth}
	        \includegraphics[width=\linewidth]{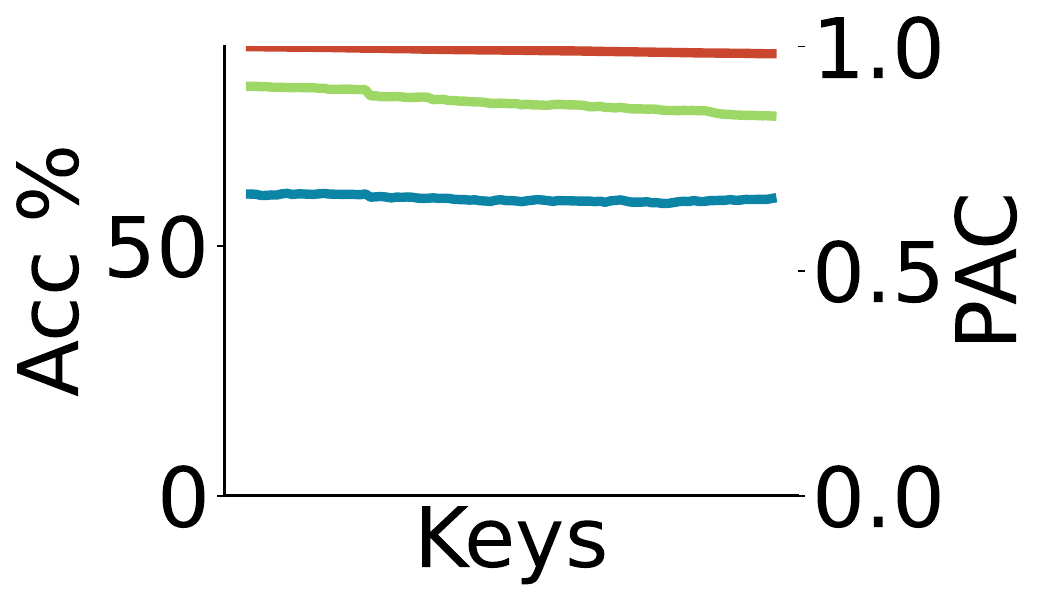}
	        \caption{Increment value}
	        \label{fig1:subfigRF}
         \end{subfigure}
	\caption{\resneteighteen~\cite{resnet} robustness protected by Hadamard/Permutation layers to key manipulation, e.g. random key, replacing of key value, and, adding noise to key. Red line depicts key similarity, Green and Blue depict train and test accuracy.}
	\label{fig:subfigures:ResNetFunctionality}
\end{figure}

To showcase the robustness of the protected \resneteighteen~\cite{resnet}, it was protected using Hadamard Protection layers and Permutation Protection layers. The results are depicted in~\cref{fig:subfigures:ResNetFunctionality}, the first row shows the results for the model protected by Hadamard layers, while the second row shows the results for the model protected by Permutation layers. For each model type three experiments identical to the ones described in~\cref{eval:subsec:generalfunc} were conducted. The first evaluation (depicted in the first column of~\cref{fig:subfigures:ResNetFunctionality}) assigned randomly generated false keys to the protected models and measured the performance and key similarity. The second experiment (second column in~\cref{fig:subfigures:ResNetFunctionality}) iteratively replaced from zero up to 100 values, within the key, with random values. The third experiment (depicted in the third column of~\cref{fig:subfigures:ResNetFunctionality} added noise to the key or incremented the key by one for the model protected using Permutation layers. All results indicate a strong robustness to key manipulation attempts similar to the CNN model.

\section{Additional Fidelity Experiments}
\begin{figure}
        \captionsetup[subfigure]{aboveskip=-2pt,belowskip=-2pt}
	\centering
	\begin{subfigure}{0.49\linewidth}
		\includegraphics[width=\linewidth]{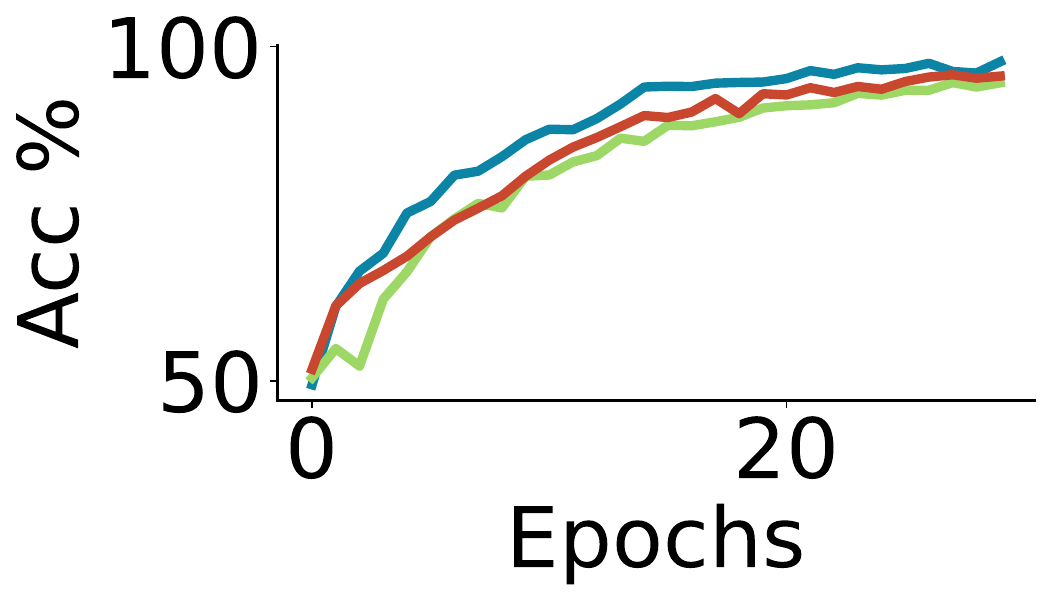}
		\caption{ResNet-18 (CIFAR-10)}
		\label{fig:subfig2:a}
	\end{subfigure}
        \hfill
	\begin{subfigure}{0.49\linewidth}
		\includegraphics[width=\linewidth]{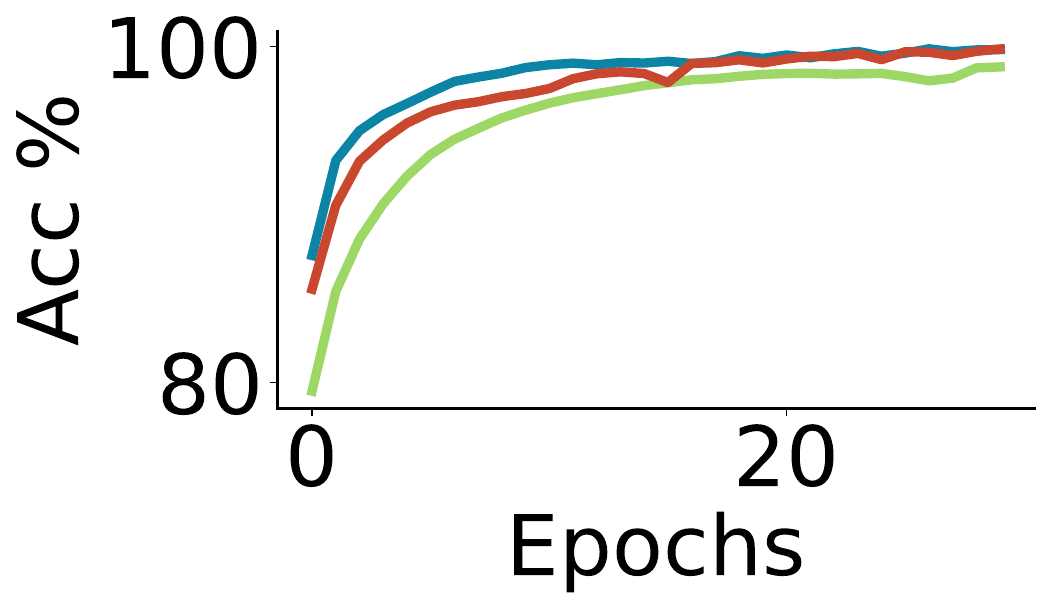}
		\caption{CNN (MNIST)}
		\label{fig:subfig2:b}
	\end{subfigure}
	\caption{The training set accuracy of three different models at each training epoch are visualized. The blue line represents the unprotected model, the red line the model protected using Permutation Layers and the green line the model protected using Hadamard layers.}
	\label{fig:baseline:hada:perm}
\end{figure}

In the following we present additional fidelity experiments for the \resneteighteen~\cite{resnet} and CNN model trained on \cifar~\cite{cifar} and \mnist~\cite{mnist}. The results, visualized in~\cref{fig:baseline:hada:perm}, depict the accuracies for the training set for the unprotected model at each training epoch as the blue line, the green line depict the model protected using Hadamard layers, and the red line the model protected using Permutation layers.

	\end{document}